\documentclass[apjl]{emulateapj}
\newcommand{\myemaila}{itoh@hep01.hepl.hiroshima-u.ac.jp}
\newcommand{\myemailb}{itoh@hp.phys.titech.ac.jp}
\usepackage{longtable}
\usepackage{graphicx}

\begin{document}

\title{Systematic Study of Gamma-ray bright Blazars
with Optical Polarization and Gamma-ray Variability}

\author{
Ryosuke Itoh\altaffilmark{1,1a},
Krzysztof Nalewajko\altaffilmark{2,2a},
Yasushi Fukazawa\altaffilmark{1},
Makoto Uemura\altaffilmark{3},
Yasuyuki T. Tanaka\altaffilmark{3},
Koji S. Kawabata\altaffilmark{3},
Grzegorz M. Madejski\altaffilmark{2},
Frank. K. Schinzel\altaffilmark{4,4a},
Yuka Kanda\altaffilmark{1},
Kensei Shiki\altaffilmark{1},
Hiroshi Akitaya\altaffilmark{3},
Miho Kawabata\altaffilmark{1},
Yuki Moritani\altaffilmark{5},
Tatsuya Nakaoka\altaffilmark{1},
Takashi Ohsugi\altaffilmark{3}
Mahito Sasada\altaffilmark{6},
Katsutoshi Takaki\altaffilmark{1},
Koji Takata\altaffilmark{1},
Takahiro Ui\altaffilmark{1},
Masayuki Yamanaka\altaffilmark{7},
and
Michitoshi Yoshida\altaffilmark{3}
}

\altaffiltext{1}{Department of Physical Science, Hiroshima
  University, Higashi-Hiroshima, Hiroshima 739-8526, Japan; \myemaila}
\altaffiltext{1a}{Department of Physics, Tokyo Institute of Technology,
  2-12-1 Ohokayama, Meguro, Tokyo 152-8551, Japan; \myemailb}
\altaffiltext{2}{Kavli Institute for Particle Astrophysics and Cosmology, 
SLAC National Accelerator Laboratory, Stanford University, 2575 Sand Hill Road M/S 29, Menlo Park, CA 94025, USA}
\altaffiltext{2a}{Nicolaus Copernicus Astronomical Center, Polish Academy of Sciences, Bartycka 18, 00-716 Warsaw, Poland}
\altaffiltext{3}{Hiroshima Astrophysical Science Center, Hiroshima
University, Higashi-Hiroshima, Hiroshima 739-8526, Japan}
\altaffiltext{4}{Department of Physics and Astronomy, University of New Mexico, Albuquerque NM, 87131, USA}
\altaffiltext{4a}{National Radio Astronomy Observatory, P.O. Box O, Socorro, NM 87801, USA}
\altaffiltext{5}{Kavli Institute for the Physics and Mathematics of the Universe (WPI),
  The University of Tokyo Institutes for Advanced Study, The University of Tokyo, Kashiwa, Chiba 277-8583, Japan}
\altaffiltext{6}{Institute for Astrophysical Research, Boston University, 725 Commonwealth Avenue, Boston, MA 02215, USA}
\altaffiltext{7}{Department of Physics, Faculty of Science and Engineering,
Konan University, Okamoto, Kobe, Hyogo 658-8501, Japan}


\begin{abstract}

Blazars are highly variable active galactic nuclei which emit radiation at all
wavelengths from radio to gamma-rays.
Polarized radiation from blazars is one key piece of evidence for
synchrotron radiation at low energies and it also varies dramatically.
The polarization of blazars is of interest for understanding the
origin, confinement, and propagation of jets.
However, even though numerous measurements have been performed, the
mechanisms behind jet creation, composition and variability are still
debated.

We performed simultaneous gamma-ray and optical photopolarimetry
observations of 45 blazars between Jul. 2008 and
Dec. 2014 to investigate the mechanisms of variability and search for
a basic relation between the several subclasses of blazars.
We identify a correlation between the maximum degree of optical linear polarization 
and the gamma-ray luminosity or the ratio of gamma-ray to optical fluxes.
Since the maximum polarization degree depends on the condition of the magnetic field (chaotic or ordered),
this result implies a systematic difference in the intrinsic alignment of magnetic fields 
in pc-scale relativistic jets between different types blazars (FSRQs vs. BL Lacs), 
and consequently between different types of radio galaxies (FR Is vs. FR IIs).

\end{abstract}

\keywords{}

\section{Introduction}

Blazars are a subclass of active galactic nuclei (AGNs) possessing relativistic
jets, which are extremely powerful and fast outflows of plasma that emerge from
the vicinity of the massive black hole.
Their observed emission is dominated by the contributions of
relativistic jets aligned with the observer's line of sight resulting in 
a strong apparent boost due to relativistic beaming.
Outstanding characteristics of blazars are their rapid and high-amplitude
intensity variations.
The apparent bolometric luminosity of blazars can be as high as 
$10^{48}$~erg~s$^{-1}$ \citep[see, e.g.,][]{1995PASP..107..803U}.  
The overall spectral energy distribution consists of at least two broad non-thermal components, the 
low-energy one attributed to synchrotron radiation, and the high-energy one attributed 
to inverse Compton scattering.
Since the non-thermal emission from jets is dominant compared
to the thermal emission from the disk due to relativistic effects,
blazars are some of the most suitable objects to study the relativistic jets.

Depending on the behavior in optical spectra or in the peak frequency 
of synchrotron radiation, 
blazars are divided into different sub-classes.
Flat spectrum radio quasars (FSRQs) are defined
to have strong emission lines of equivalent width $> 5\ $\AA\ in the observer's optical 
band \citep{1991ApJ...374..431S}.  In contrast, BL Lac objects show relatively weak emission lines.
Additionally, blazars can be classified into three types based on their peak frequency 
of synchrotron radiation $\nu_{\rm peak}^{S}$:  low-synchrotron-peaked blazars (LSP; for 
sources with $\nu_{\rm peak}^{S} < 10^{14}$ Hz), intermediate-synchrotron-peaked blazars (ISP; 
for $10^{14}\;{\rm Hz} < \nu_{\rm peak}^{S} < 10^{15}$ Hz), and high-synchrotron-peaked blazars 
(HSP; for $10^{15}\;{\rm Hz} < \nu_{\rm peak}^{S}$) \citep{2011ApJ...743..171A,2015ApJ...810...14A}.  
Essentially most of FSRQs are LSPs 
\citep[only three FSRQ-HSP and several FSRQ-ISP are reported in][]{2015ApJ...810...14A}, 
while BL Lacs can be LSPs, ISPs or HSPs.  
According to the blazar sequence, the synchrotron luminosity $L_{\rm syn}$, the inverse Compton luminosity $L_{\rm IC}$, 
and also the ratio of these two luminosities $q = L_{\rm IC}/L_{\rm syn}$ are inversely correlated 
with the synchrotron peak frequency $\nu_{\rm peak}^{S}$ \citep{1998MNRAS.299..433F}.
Hence, the FSRQs are both more luminous and more Compton dominated (higher $q$ value) than the BL Lacs.
\cite{1998MNRAS.301..451G} provided a physical explanation for this spectral sequence,
proposing that it originated from radiative electron cooling in the jet.
A unified model of \cite{1995PASP..107..803U} has become generally accepted, whereas FSRQs are related 
to intrinsically powerful (FR II) radio galaxies, and BL Lac objects are related to intrinsically 
weak (FR I) radio galaxies. 
The two types of radio galaxies also possess jets, 
but those are directed farther away from our line of sight.  

\cite{1990A&AS...83..183M} performed a large-sample study of blazars in the optical band 
and showed that high polarization degree and variability of polarization are common phenomena 
in blazars:  this, together with the high level of polarization observed in the radio 
band, provides strong support for the synchrotron radiation as an origin of the low energy emission.  
The level, but also the position angle of polarization 
(electric vector position angle, the direction is measured from north to east) in blazars often 
varies dramatically, and these are important ingredients for understanding the origin, 
confinement, and propagation of jets \citep[e.g.,][]{1985agn..book..215B,1998AJ....116.2119V}.
\citet[][hereafter Paper I]{2011PASJ...63..639I}, 
reported the statistics of photopolarimetric observations of 
blazars on daily timescales, and suggested that sources characterized by lower luminosity, 
and those with the peak of the synchrotron radiation located at higher frequencies 
(such as HSPs) had smaller amplitude variations in the flux, color, 
and polarization degree.  These authors also reported that about 30\% of blazars showed 
a correlation between the optical flux and polarization degree.
Polarization depends on the structure of the magnetic field in the emitting region,
and thus polarimetric observations of blazars in the optical band are valuable 
for probing the magnetic fields in relativistic AGN jets at (sub-)pc scales
since the optical emission region is thought to be located at pc-scales from central engine 
\citep[e.g.,][]{2008Natur.452..966M,2011ApJ...735L..10A}.
Rotations of polarization angle during flares are also important observational phenomena in blazars
\citep[e.g.][]{2008Natur.452..966M,2015MNRAS.453.1669B}.
However, the details need to be dealt with carefully, because 
some apparent rotations might be caused by random variation of polarization on the Stokes parameter QU plane
\citep[e.g.][]{1985ApJ...290..627J,2016A&A...590A..10K}.

Very few attempts have been made to systematically study variability, 
especially focusing on multi-wavelength and polarimetric observations amongst 
the several subclasses of blazars \citep{2015MNRAS.453.1669B}.
In this paper, we search for a basic relation between gamma-ray properties and 
optical flux and polarization with a systematic study of 45 blazars to 
investigate the mechanisms of variability.

\section{Observations}

\subsection{Optical Observations with Kanata}

We performed optical and near infrared imaging polarimetry of 42 AGNs between
Aug. 2008 and Dec. 2014 with the 1.5m diameter Kanata Telescope.  We used two instruments 
attached to the Kanata telescope:  
one is TRISPEC \citep[Triple Range Imager and SPECtrograph;][]{2005PASP..117..870W}
and the other is HOWPol \citep[Hiroshima One-shot Wide-field Polarimeter;][]{2008SPIE.7014E.151K}.
TRISPEC was attached to the Cassegrain focus of the Kanata telescope from 2006 to 2011 and
it has a CCD and two InSb arrays, enabling photopolarimetric observations
in one optical and two near-infrared bands simultaneously.
HOWPol is installed at the Nasmyth focus of the Kanata telescope, and has been in operation
since 2009.

We performed {\it V}, {\it J}, {\it Ks}-band photometry and polarimetry
observations of each target from July 2008 to February 2010 using
TRISPEC and performed the {\it V} and {\it R$_C$}-band photometry and
polarimetry observations from July 2008 to February 2010 using HOWPol.
Each observing sequence consisted of successive exposures at four position angles of
a half-wave plate of $0^{\circ}, 45^{\circ}, 22.5^{\circ}$ and $67.5^{\circ}$.

The data reduction involved standard CCD photometry procedures ---
aperture photometry using \verb|APPHOT| package in \verb|PYRAF| 
and differential photometry with a comparison star taken in the same frame.
The positions of the comparison stars are listed in Table \ref{tab:KanataObs1}.
The data have been corrected for Galactic extinction (values are given in Table \ref{tab:KanataObs1}).
We confirmed that the instrumental polarization was smaller than 0.1\%
in the V band (TRISPEC), using unpolarized standard stars and thus applied no
correction for it. The polarization angle (PA) is defined in the standard manner 
(measured from north to east), and it was calibrated with two polarized stars, 
HD19820 and HD25443 \citep{1996AJ....111..856W}.
The time series data of {\it V},  {\it R$_C$}, {\it J}, {\it Ks}-band photometry and
{\it V}, {\it R$_C$}-band polarimetry will be available via
Centre de Donn\'ees astronomiques de Strasbourg (CDS, Strasbourg astronomical Data Center
\footnote{http://cds.u-strasbg.fr/}).

Polarimetry with HOWPol suffers from large instrumental polarization ($\Delta p\sim 4$\%) 
produced by the reflection of the incident light on the tertiary mirror of the telescope.
The instrumental polarization was modeled as a function of the declination of the object and the 
hour angle at the observation, and we subtracted it from the observation.
We estimated that the error in this instrumental polarization correction is smaller 
than $\Delta p\sim0.5$\% from many observations of unpolarized standard stars.
The PA was calibrated using two polarized stars, HD183143 and HD204827
\citep{1983A&A...121..158S}.
We also confirmed that systematic differences in the photometric and polarimetric 
systems are negligibly small by measurements of comparison stars.

\begin{table*}
  \begin{center}
  \caption{List of comparison stars}
  \label{tab:KanataObs1}  
  \begin{tabular}{lccccccc}
    \hline\hline
    Source Name    & Comparison Coordinates & $V$    & $R_C$  & $J$    & $Ks$   & A(V)  & Ref.    \\ 
    (1)            & (2)                    & (3)    & (4)    & (5)    & (6)    & (7)   & (8)     \\ \hline
    PKS 0048-097   & 00:50:47.0 -09:30:15.0 & 14.096 & 13.741 & 12.455 & 11.854 & 0.104 & [1],[2] \\ 
    S2 0109+22     & 01:12:03.0 +22:43:26.0 & 12.477 & 12.272 & 11.245 & 10.886 & 0.122 & [1],[2] \\
    Mis V1436      & 01:36:42.0 +47:51:03.0 & 13.394 & 13.272 & 12.223 & 11.922 & 0.496 & [1],[2] \\
    PKS 0215+015   & 02:17:49.0 +01:48:28.0 & 12.156 & 12.076 & 11.320 & 11.046 & 0.108 & [1],[2] \\
    3C 66A         & 02:22:55.1 +43:03:15.5 & 13.183 & 13.314 & 12.371 & 12.282 & 0.274 & [3],[4] \\
    AO 0235+164    & 02:38:32.0 +16:36:00.0 & 12.756 & 12.523 & 11.248 & 10.711 & 0.258 & [1],[2] \\
    1H 0323+342    & 03:24:39.0 +34:11:29.0 & 13.445 & 12.773 & --     & --     & 0.680 & [1]     \\
    \              & 03:24:33.0 +34:10:53.0 & 13.221 & --     & 11.232 & 10.589 & 0.680 & [3],[2] \\
    1ES 0323+022   & 03:26:13.4 +02:24:06.1 & 12.840 & --     & 11.097 & 10.485 & 0.551 & [5]     \\
    PKS 0422+00    & 04:24:42.4 +00:37:10.8 & 12.510 & --     & 11.217 & 10.899 & 0.338 & [5]     \\
    PKS 0454-234   & 04:57:00.0 -23:26:05.0 & 12.359 & 12.149 & 10.849 & 10.364 & 0.149 & [1],[2] \\
    1ES 0647+250   & 06:50:40.0 +25:03:24.0 & 13.060 & 12.921 & 12.053 & 11.771 & 0.320 & [1],[2] \\
    S5 0716+714    & 07:21:54.0 +71:19:20.0 & 13.468 & 13.367 & --     & --     & 0.101 & [1]     \\
    & 07:21:52.2 +71:18:16.1 & 12.345 & --     & 11.320 & 10.980 & 0.101 & [3],[4] \\
    4C 14.23       & 07:25:20.0 +14:25:03.0 & 14.842 & 14.637 & --     & --     & 0.284 & [1]     \\
    PKS 0735+17    & 07:38:02.0 +17:41:21.0 & 14.170 & 13.991 & --     & --     & 0.110 & [1]     \\
    PKS 0754+100   & 07:57:16.1 +09:55:47.8 & 13.000 & --     & 11.852 & 11.496 & 0.071 & [4]     \\
    1ES 0806+524   & 08:09:40.0 +52:19:17.0 & 12.999 & 12.741 & 11.417 & 10.867 & 0.146 & [1],[5] \\
    OJ 49          & 08:31:54.0 +04:30:43.0 & 13.550 & 13.429 & --     & ---    & 0.106 & [1]     \\
    & 08:32:00.7 +04:32:02.5 & 13.517 & --     & 12.475 & 12.189 & 0.106 & [3],[5] \\
    OJ 287         & 08:54:53.0 +20:04:44.0 & 14.190 & 13.929 & --     & --     & 0.092 & [1]     \\
    & 08:54:59.0 +20:02:57.1 & 13.954 & 13.832 & 12.811 & 12.445 & 0.092 & [3],[4] \\
    PMN J0948+022  & 09:49:10.0 +00:21:40.0 & 14.983 & 14.732 & 13.500 & 13.139 & 0.253 & [1],[2] \\
    S4 0954+65     & 09:58:50.4 +65:32:09.1 & 14.610 & --     & 12.927 & 12.455 & 0.372 & [5]     \\
    Mrk 421        & 11:04:18.2 +38:16:30.5 & 15.570 & 15.200 & 14.453 & 14.106 & 0.050 & [6]     \\
    ON 325         & 12:17:44.0 +30:09:43.0 & 15.097 & 14.871 & 13.674 & 13.232 & 0.075 & [1],[2] \\
    1ES 1218+304   & 12:21:31.0 +30:11:00.0 & 12.400 & 10.489 & --     & --     & 3.240 & [7]     \\
    ON 231         & 12:21:33.0 +28:13:04.0 & 12.071 & 11.965 & 10.921 & 10.597 & 0.076 & [1],[2] \\
    3C 273         & 12:29:08.0 +02:00:18.0 & 12.725 & 12.540 & 11.345 & 10.924 & 0.067 & [1],[4] \\
    GB6 J1239+0443 & 12:39:30.1 +04:39:52.6 & 14.095 & --     & 12.942 & 12.638 & 0.072 & [8]     \\
    3C 279         & 12:56:10.0 -05:50:14.0 & 12.420 & 12.257 & --     & --     & 0.093 & [1]     \\
    & 12:56:16.9 -05:50:43.0 & 13.517 & 13.318 & 12.377 & 11.974 & 0.093 & [3],[9] \\
    OQ 530         & 14:20:18.0 +54:24:14.0 & 14.357 & 14.189 & --     & --     & 0.043 & [1]     \\
    & 14:19:39.7 +54:21:55.0 & 16.009 & --     & 13.873 & 13.131 & 0.043 & [3],[8] \\
    PKS 1502+106   & 15:04:13.0 +10:28:42.0 & 14.552 & 14.331 & --     & --     & 0.104 & [1]     \\
    & 15:04:36.5 +10:28:47.0 & 15.328 & --     & 14.117 & 13.678 & 0.104 & [3],[5] \\
    PKS 1510-089   & 15:12:51.0 -09:05:23.0 & 14.630 & 14.466 & --     & --     & 0.327 & [1]     \\
    & 15:12:53.2 -09:03:43.6 & 13.195 & --     & 12.205 & 11.919 & 0.327 & [3],[4] \\
    RX J1542.8+612 & 15:42:39.0 +61:30:26.0 & 13.958 & 13.303 & 9.640  & --     & 0.052 & [1],[4] \\
    PG 1553+113    & 15:55:52.0 +11:13:18.0 & 13.842 & 13.625 & 12.539 & 12.139 & 0.169 & [1],[9] \\
    3C 345         & 16:42:52.0 +39:48:33.0 & 15.304 & 14.963 & --     & --     & 0.043 & [1]     \\
    Mrk 501        & 16:53:45.0 +39:44:09.0 & 12.534 & 12.195 & 10.935 & 10.399 & 0.061 & [1],[4] \\
    H1722+119      & 17:25:05.0 +11:52:10.0 & 13.214 & 12.828 & 11.308 & 10.710 & 0.559 & [1],[5] \\
    NRAO 530       & 17:33:00.0 -13:04:09.0 & 14.488 & 13.851 & --     & --     & 1.699 & [1]     \\
    PKS 1749+096   & 17:51:31.0 +09:39:40.0 & 14.278 & 14.014 & --     & --     & 0.577 & [1]     \\
    & 17:51:37.3 +09:39:07.1 & 11.857 & --     & 10.252 & 9.740  & 0.577 & [3],[5] \\
    S5 1803+784    & 17:59:52.6 +78:28:50.9 & 13.133 & 12.226 & 11.761 & 11.381 & 0.169 & [7],[10]\\
    3C 371         & 18:07:12.0 +69:47:07.0 & 14.127 & 13.900 & --     & --     & 0.109 & [1]     \\
    & 18:06:53.7 +69:45:37.4 & 13.254 & --     & 12.219 & 11.856 & 0.109 & [3],[4] \\
    1ES 1959+650   & 19:59:39.2 +65:08:52.9 & 14.618 & 11.301 & --     & --     & 0.557 & [7]     \\
    & 20:00:26.5 +65:09:26.4 & 13.180 & --     & 11.464 & 11.315 & 0.557 & [6]     \\
    PKS 2155-304   & 21:59:02.5 -30:10:46.2 & 12.050 & --     & 10.775 & 10.365 & 0.071 & [4]     \\
    BL Lac         & 22:02:45.0 +42:16:35.0 & 12.939 & 12.326 &  9.817 &  8.811 & 1.063 & [1],[4] \\
    CTA 102        & 22:32:41.0 +11:43:14.0 & 15.347 & 14.971 & --     & --     & 0.233 & [1]     \\    
    3C 454.3       & 22:53:58.0 +16:09:06.0 & 13.661 & 13.342 & 11.858 & 11.241 & 0.349 & [1],[4] \\
    1ES 2344+514   & 23:47:02.2 +51:43:17.6 & 12.565 & 12.177 & 11.421 & 11.117 & 0.680 & [7],[5] \\
    \hline \\
  \end{tabular}
  \end{center}

  {\footnotesize (1) Object name. (2) Coordinate of comparison stars. 
    (3), (4), (5), (6) V, R, J, Ks band magnitudes of comparison stars.
    (7) Galactic extinction for V-band.
    (8) Reference for the magnitudes of comparison stars. [1];UCAC-4 catalog, [2];2MASS catalog, 
    [3];Calibrated with UCAC-4, [4]Gonzalez-Perez+01, [5];Skiff+05, 
    [6];Villata+98, [7];UCAC-3, [8];Adelman-McCarthy+07, [9];Doroshenko+05, [10];Zacharias+05}

\end{table*}

\subsection{Gamma-ray Observations with Fermi}

The {\it Fermi} Gamma-ray Space Telescope is an observatory in a low-Earth 
orbit launched on 2008 June 11.
The Large Area Telescope (LAT) 
is the instrument used for monitoring high-energy (MeV to GeV) emission of
AGN and other sources.
It is an electron-positron pair 
production detector with a bandpass of 20 MeV - 300 GeV, 
described in detail in \cite{2009ApJ...697.1071A}.  
The LAT observes the whole sky every 3 hours with a large 
effective area of $8000$ cm$^2$ at 1 GeV, a wide field of view of 
$2.4$ sr, and a single photon angular resolution 
(68\% containment radius) of $0.6^{\circ}$ at 1 GeV.

The data used in this analysis were taken between 2008
August and 2014 December, almost entirely in sky survey mode.  
The data were analyzed using the standard \emph{Fermi} analysis software
(\verb|Science Tools|, version v10r00, IRFs P8$\_$R2).  We use Pass 8 ``Source'' 
class event data above 100 MeV.  We also restricted our analysis to events 
with zenith angles $<90^{\circ}$ to limit the contamination by gamma-rays 
from the Earth's limb.  We performed an {\it Unbinned Likelihood} analysis to calculate the
gamma-ray spectrum and flux of our targets, using the \verb|gtlike|
package in the \verb|Science Tools|.  An area of $15^{\circ}$ around target was 
selected as a region of interest (ROI) for this analysis.  We constructed 
a model of the ROI that includes a point source at the position of each target.
We modeled the spectrum of each blazar as a power-law: 
\begin{equation}
\frac{dN}{dE} = N_0 \left(\frac{E}{E_0}\right)^{-\alpha}  \label{eq:pl}
\end{equation}
where $N_0$ is the normalization at energy $E_0$ and $\alpha$ is the
photon index.
The flux normalization and spectral index were left free in the likelihood analysis.
We constructed the background source model based on the 3FGL catalog 
\citep[][LAT 4-year Point Source Catalog]{2015ApJS..218...23A}.
The spectral indicies of the background sources were fixed to their catalog values while their
normalizations were left free.
The Galactic diffuse emission component \citep[{\tt gll\_iem\_v06.fit},][]{2016ApJS..223...26A}
and the isotropic diffuse emission component ({\tt iso\_P8R2\_SOURCE\_V6\_v06.txt})
are included in our models.
We performed the model fitting in twice. 
First, all the 3FGL sources in the ROI were included in the model and
fitted over the 7-day intervals.
We then fit a second model with the background sources with low test
statistics ($TS < 25$) omitted from the data.

\subsection{Data selection for systematic studies}
\label{sec:ana}
The purpose of this study is to search for a basic relation between the blazars sub classes.
Both gamma-ray and optical band data possess observational gaps due to 
low photon statistics, bad weather, visibility and maintenance of instruments.
Some parameters, like the variability index (see below for the definition),
are dependent on their observational periods.
In order to compare such parameters between the gamma-ray and optical band,
we selected strictly simultaneous data from our sample.  Specifically, 
we excluded the gamma-ray data which have no counterpart in optical observations.  
We also extracted the optical data which have no significant gamma-ray detection 
($TS < 25$).
Since the bin size of the gamma-ray data was set to 7 days, 
we also averaged the optical data to match the
gamma-ray bins for the DCF analysis (Section \ref{sec:DCF}) and
the derivation of the ratio between gamma-ray flux and optical flux.
The unaveraged data were used in all other cases.
We note that there are several 
uncertainties in the calculation of the time lag since 
averaging the optical data may make some temporal bias
since the observational time of gamma-ray does not necessarily correspond
to average time of optical data.
Thus, we do not discuss time lags less than 3.5 days in this paper.
Among 45 blazars, we selected 24 targets for our systematic studies,
which have more than 10 simultaneous gamma-ray and optical data points.  
These targets are listed in Table \ref{tab:24sources}.

\begin{table*}
  \begin{center}
  \caption{List of our targets with more than 10 data points.}
  \label{tab:24sources}
  \begin{tabular}{lcccccc}\hline\hline
    Object Name   & 3FGL name         & $\log(v_{\rm peak})$ & Type        & z     & $N_{\rm opt.} $ & $N_{\rm \gamma} $  \\ 
    (1)           & (2)               & (3)               & (4)         & (5)   & (6)          & (7)  \\ \hline
    S2 0109+22    & 3FGL J0112.1+2245 & 14.6              & ISP         & 0.265 & 44  & 24  \\
    Mis V1436     & 3FGL J0136.9+4751 & 13.6              & LSP (FSRQ)  & 0.859 & 52  & 18  \\
    3C 66A        & 3FGL J0222.6+4302 & 15.1              & ISP         & 0.444 & 462 & 164 \\
    AO 0235+164   & 3FGL J0238.7+1637 & 13.5              & LSP         & 0.94  & 72  & 26  \\
    PKS 0454-234  & 3FGL J0457.0-2325 & 13.1              & LSP (FSRQ)  & 1.003 & 27  & 20  \\
    S5 0716+714   & 3FGL J0721.9+7120 & 14.6              & ISP         & 0.3   & 556 & 198 \\
    OJ 49         & 3FGL J0831.9+0429 & 13.5              & LSP         & 0.1737& 27  & 16  \\
    OJ 287        & 3FGL J0854.8+2005 & 13.4              & LSP         & 0.306 & 174 & 75  \\
    Mrk 421       & 3FGL J1104.4+3812 & 16.6              & HSP         & 0.031 & 85  & 46  \\
    ON 325        & 3FGL J1217.8+3006 & 15.5              & HSP         & 0.13  & 38  & 17  \\
    3C 273        & 3FGL J1229.1+0202 & 13.5              & LSP (FSRQ)  & 0.15834& 224& 91  \\
    3C 279        & 3FGL J1256.1-0547 & 12.6              & LSP (FSRQ)  & 0.5362& 140 & 72  \\
    PKS 1502+106  & 3FGL J1504.3+1029 & 13.6              & LSP (FSRQ)  & 1.839 & 71  & 27  \\
    PKS 1510-089  & 3FGL J1512.8-0906 & 13.1              & LSP (FSRQ)  & 0.36  & 108 & 51  \\
    RX J1542.8+612& 3FGL J1542.9+6129 & 14.1              & LSP (FSRQ)  & 0.117 & 69  & 38  \\
    PG 1553+113   & 3FGL J1555.7+1111 & 15.4              & HSP         & 0.36  & 196 & 90  \\
    Mrk 501       & 3FGL J1653.9+3945 & 17.1              & HSP         & 0.033663&170& 80  \\
    PKS 1749+096  & 3FGL J1751.5+0938 & 13.1              & LSP (FSRQ)  & 0.322 & 47  & 16  \\
    3C 371        & 3FGL J1806.7+6948 & 14.7              & ISP (FSRQ)  & 0.051 & 21  & 16  \\
    1ES 1959+650  & 3FGL J2000.0+6509 & 16.6              & ISP         & 0.047 & 82  & 42  \\
    PKS 2155-304  & 3FGL J2158.8-3013 & 16.0              & HSP         & 0.116 & 146 & 60  \\
    BL Lac        & 3FGL J2202.8+4216 & 13.6              & LSP         & 0.0686& 340 & 137 \\
    CTA 102       & 3FGL J2232.4+1143 & 13.6              & LSP (FSRQ)  & 1.07  & 76  & 33  \\    
    3C 454.3      & 3FGL J2253.9+1609 & 13.6              & LSP (FSRQ)  & 0.859 & 442 & 143 \\
    \hline
  \end{tabular}
  \end{center}
      {\footnotesize (1) Object name. (2) object name in 3FGL catalog. (3) Synchrotron peak frequency. (4) blazar type. (5) redshift of object from the NASA/IPAC Extragalactic Database (NED). 
        (6) Number of simultaneous optical ($V$-band) data point. (7) Number of simultaneous GeV data point.}
\end{table*}

\section{Results}
\label{sec:res}

There are a lot of evaluation methods of blazar properties,
such as measurement of variability index, correlation between gamma-ray
and optical properties.
In addition to these, blazars have various characteristics such as luminosity,
redshift and synchrotron peak frequency.
The question we have to ask here is what are common properties in blazars.
In this section, we describe results with several blazar properties 
which might be good elements of a new blazar classification scheme.

\subsection{Light curves}

\begin{figure}
  \centering
  \includegraphics[angle=0,width=8cm]{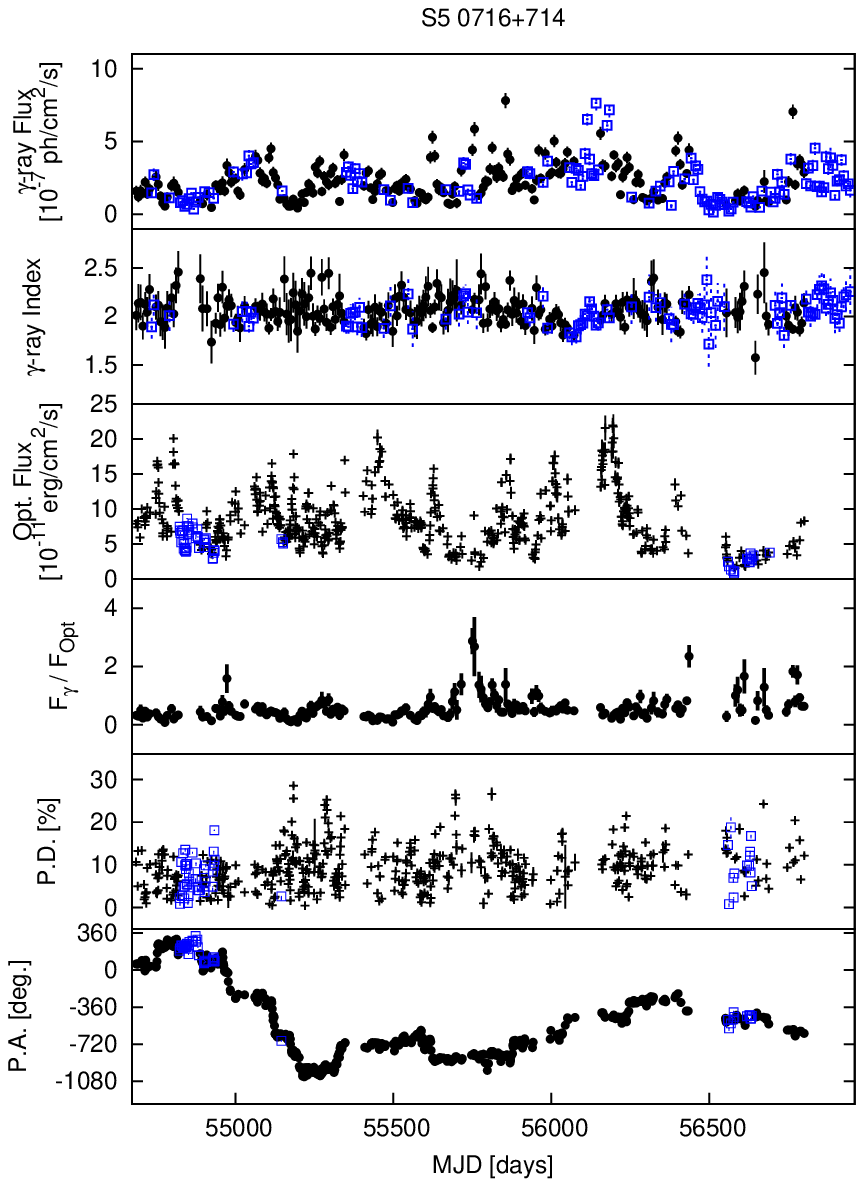}
  \caption{Temporal variability of gamma-ray flux, gamma-ray index, optical flux,
    ratio of gamma-ray flux, polarization degree and polarization angle for the source S5~0716+714.
    The top panel shows the gamma-ray light curve, 
    the second panel shows variability of gamma-ray spectral index,
    the third panel shows the optical V-band light curve,
    the fourth panel shows ratio of gamma-ray to optical flux,
    the fifth panel shows variability of polarization degree and
    bottom panel shows variability of polarization angle. 
    Black-filled data points indicates the data with simultaneous optical polarization and gamma-ray observations that we used in our analysis.
    Blue-opened box data points indicate the excluded data in our analysis
    (see Sec \ref{sec:ana} for data selection).}
  \label{fig:MWLC_s5_0716}
\end{figure}

In this section, we report on the results of our observations.
In Figure \ref{fig:MWLC_s5_0716},
we show the temporal variation of the gamma-ray flux, gamma-ray index,
optical flux, ratio of gamma-ray to optical fluxes, 
polarization degree, and polarization angle for the source S5~0716+714.
The ratio of gamma-ray to optical fluxes is derived from the gamma-ray 
$\nu F_{\nu}$ value at 100 MeV and from the optical $\nu F_{\nu}$ value in the V-band.
This ratio is similar to the Compton dominance \citep[e.g.,][]{2013ApJ...763..134F}, 
although an accurate value of the Compton dominance should be calculated from the 
integrated fluxes of synchrotron and inverse Compton components.  
This ratio can be a good indicator of conditions in the blazar jet, 
because it is only weakly redshift dependent.
Since both the gamma-ray and optical fluxes show dramatic variability,
observation of both fluxes should be simultaneous as much as possible.
In this paper, we present data for a number of blazars;  
analogous plots for all 24 objects, for completeness
including S5 0716+714, are shown in Appendix A.

There are several types of variability in different bands.
For example, S5 0716+714 shows many high-amplitude flares at very short intervals
\cite[this is also suggested in][]{2008PASJ...60L..37S}.  
In contrast, in the case of PKS 1510-089 we rarely observe very prominent flares,
which consist of a small number of subcomponents (see Fig. 21).
The cadence and the amplitudes of flares in PKS~1510-089 are also different from those in S5~0716+714.
In order to compare the properties of such different types of flares,
we investigate the observed photo-polarimetric variability in several ways.

\subsection{Variability of polarization in the Stokes parameter (Q,U) plane}

\begin{figure*}
  \centering
  \includegraphics[angle=0,width=8cm]{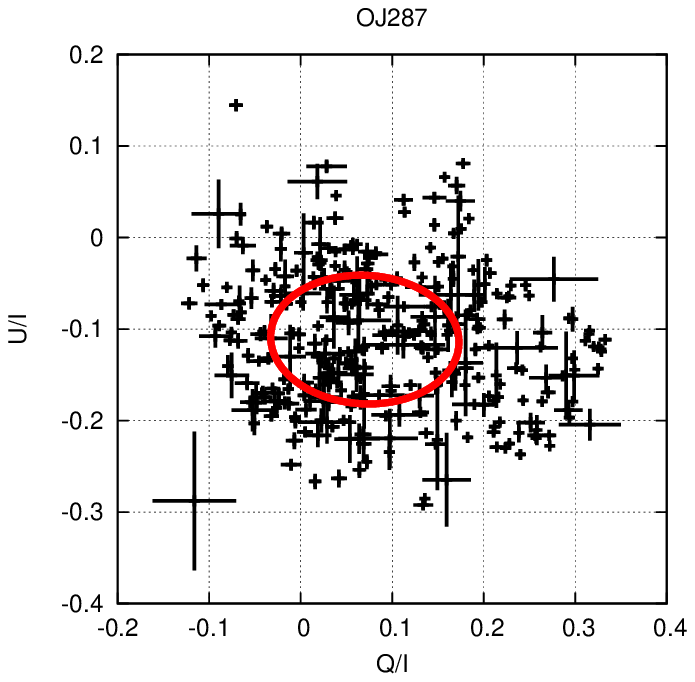}
  \includegraphics[angle=0,width=8cm]{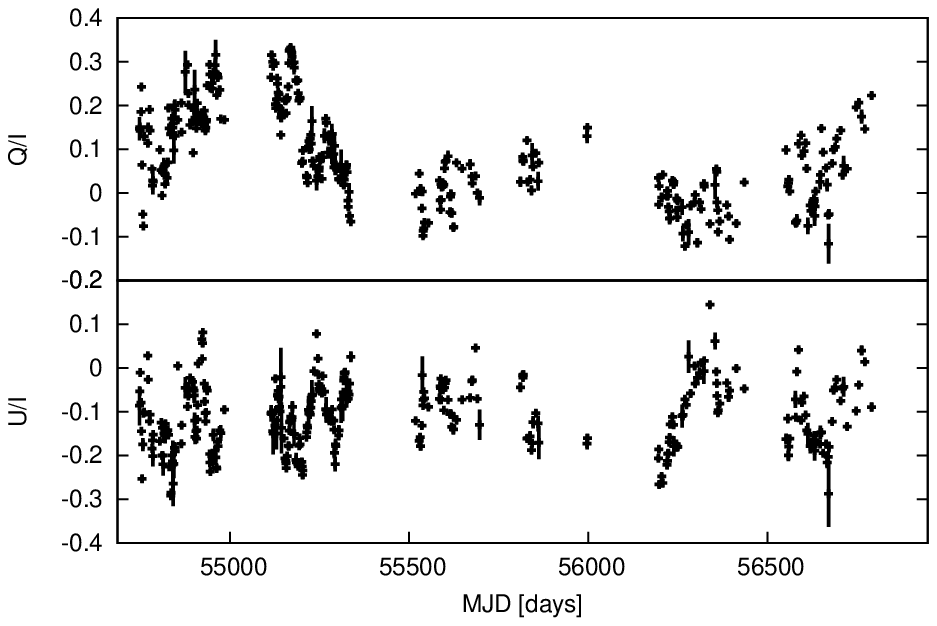}
  \caption{Left panel; Elliptical fitting in the QU plane for the source OJ 287.
    Length of major and minor axes of distribution indicates the variation (1-$\sigma$) of 
    polarization in two-dimensional space.
    Right panel; temporal variability of Q/I and U/I plot for the same source. 
    }
  \label{fig:QUellipse}
\end{figure*}

Information on linear polarization can be represented in several ways. 
Polarization degree $\Pi$ can be combined with the total flux $I$ to yield the polarized flux $\Pi I$. 
Polarization angle $\chi$ can be combined with the polarized flux to yield the Stokes parameters 
$Q = \Pi I \cos(2\chi)$ and $U = \Pi I \sin(2\chi)$.
Namely, when plotted on the ($Q/I, U/I$) plane, the distance of an 
individual point from the origin corresponds to $\Pi$, and the
direction of ($Q/I, U/I$) vector from $Q/I$ axis correspond to $2\chi$.
In some cases, blazars show a clear correlation between total optical flux 
and polarization degree, 
but sometimes they do not show such correlation. 
It is known that variability of polarization of blazars does not show simple, symmetrical motion
but shows complex motion in the $(Q,U)$ plane \citep[e.g.][]{2010PASJ...62...69U}.
In addition, polarization angle has an ambiguity of $\pm 180n$ degree (where $n$ is an integer) and this makes it difficult to measure the variability of polarization angle.
Therefore, it is important to evaluate the variability of polarization in the $(Q,U)$ plane.
The purpose of this analysis is to clarify the distribution of polarization in the $(Q,U)$ plane and
uniformity of polarization variability.
In order to compare the polarization variability properties between individual sources,
we adopted an ellipsoidal variance measurement to characterize the distribution 
of the observed polarization in the $(Q,U)$ plane.  
First, we calculated the median values of $Q_0$ and $U_0$ in order to determine the slope of major 
axis of the distribution, then we performed a two-dimensional 
least-squares fit to the observed values of $Q-Q_0$ and $U-U_0$, 
which yields the correlation coefficient, the mean polarization angle $\chi_0$ 
corresponding to the inclination of the major axis of the distribution, and also the 
variance values measured along the major and minor axes.
An example of such ellipsoidal variance for the source OJ 287 is shown in Figure \ref{fig:QUellipse}.
From this figure, one can see that the average values of $Q_0$ and $U_0$ are clearly not 
coincident with the origin of 
the $(Q,U)$ plane, and that the distribution of $Q$ and $U$ is asymmetric with respect to point ($Q_0, U_0$).
A summary of ellipsoidal variance measurement results for all sources is presented in Table \ref{tab:QUellipse}.

\begin{table}
  \centering
  \caption{Summary of variability of polarization on the Stokes parameter QU plane}
  \label{tab:QUellipse}
  \begin{tabular}{lrrrrr}\hline\hline
    Source Name   & $Q_0^{(1)}$ & $U_0^{(2)}$ & $\chi_0^{(3)}$ [deg.] &$\sigma_{\rm major}^{(4)}$ & $\sigma_{\rm minor}^{(5)}$ \\ \hline
    S2 0109+22    &  -0.04 &  -0.03 &  0.4  & 0.06 & 0.10 \\
    Mis V1436     &  -0.12 &  0.05  &  1.4  & 0.15 & 0.07 \\
    3C 66A        &  0.07  &  0.07  &  -0.2 & 0.04 & 0.05 \\
    AO 0235+164   &  -0.08 &   0.00 &  -14.5& 0.11 & 0.08 \\
    PKS 0454-234  &  0.02  &  -0.03 &  25.9 & 0.11 & 0.10 \\
    S5 0716+714   &  -0.02 &  0.02  &  -1.1 & 0.07 & 0.07 \\
    OJ 49         &  -0.03 &  -0.03 &  2.5  & 0.05 & 0.06 \\
    OJ 287        &  0.06  &  -0.11 &  -2.6 & 0.10 & 0.07 \\
    Mrk 421       &  0.01  &  -0.01 &  3.2  & 0.02 & 0.02 \\
    ON 325        &  0.07  &  -0.03 &  -33.5& 0.04 & 0.03 \\
    3C 273        &   0.00 &  0.00  &  1.0  & 0.00 & 0.00 \\
    3C 279        &  -0.06 &  0.08  &  -9.9 & 0.10 & 0.11 \\
    PKS 1502+106  &  -0.04 &  -0.15 &  -37.5& 0.16 & 0.11 \\
    PKS 1510-089  &  0.01  &  0.00  &  13.1 & 0.05 & 0.07 \\
    RX J1542.8+612&  0.03  &  0.02  &  -20.2& 0.05 & 0.04 \\
    PG 1553+113   &  -0.01 &  -0.01 &  2.7  & 0.02 & 0.03 \\
    Mrk 501       &   0.00 &  -0.01 &  -13.7& 0.01 & 0.01 \\
    PKS 1749+096  &  -0.01 &  -0.01 &  34.6 & 0.10 & 0.09 \\
    3C 371        &  -0.06 &  0.02  &  37.0 & 0.03 & 0.02 \\
    1ES 1959+650  &  0.02  &  -0.03 &  -6.8 & 0.02 & 0.02 \\
    PKS 2155-304  &  -0.02 &  0.02  &  -22.5& 0.03 & 0.03 \\
    BL Lac        &  0.07  &  0.04  &  -3.6 & 0.06 & 0.05 \\
    CTA 102       &  0.00  &  0.03  &  10.7 & 0.06 & 0.08 \\    
    3C 454.3      &  0.01  &  -0.01 &  -3.7 & 0.06 & 0.05 \\
    \hline
  \end{tabular}
  \\ (1),(2); Median value of $Q$ and $U$, 
  (3); mean polarization angle, (4),(5) the variance
values measured along the major and minor axes.
\end{table}

\subsection{Correlation between gamma-ray and optical light curves}
\label{sec:DCF} 

\begin{figure}
  \centering
  \includegraphics[angle=0,width=10cm]{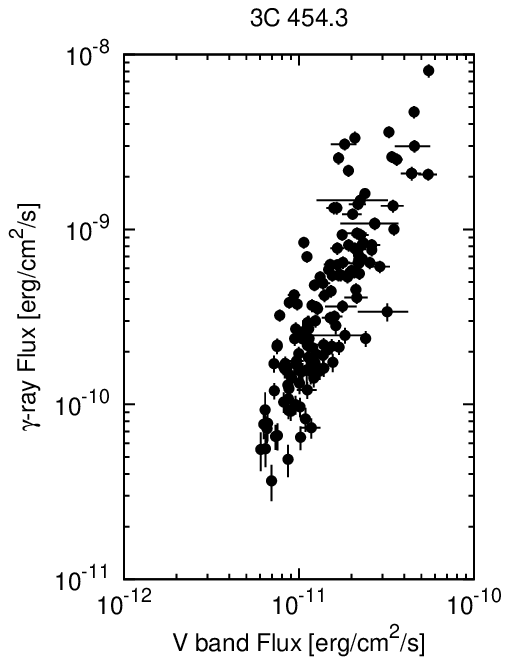}
  \caption{Scatter plot of gamma-ray flux and optical flux for 3C~454.3.
    The optical data were averaged with 7-day intervals. 
    Error bars for the optical data indicate the 1-$\sigma$ variance of 7-day interval data.
    The Figure shows significant correlation between gamma-ray flux and optical flux with no time lag
    (also see Figure \ref{fig:dcf} and Table \ref{tab:dcf}).}
  \label{fig:Sc_3c454}
\end{figure}

\begin{figure}
  \centering
  \includegraphics[angle=0,width=8cm]{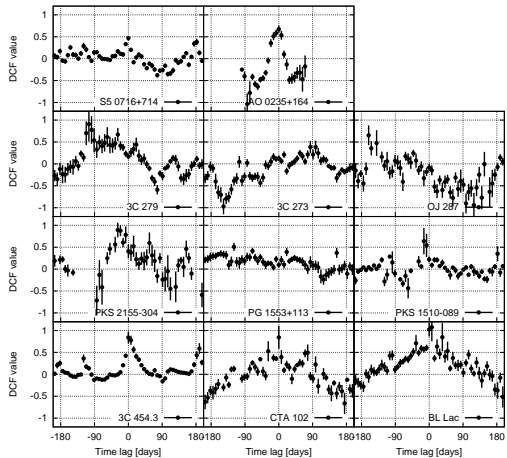}
  \caption{DCF plot of gamma-ray flux and optical flux for several blazars.
  Significant peak of each source are listed in Table \ref{tab:dcf}.}
  \label{fig:dcf}
\end{figure}

Temporal correlations between various gamma-ray and optical properties
can be quantified by calculating the Discrete Correlation Function
\citep[][DCF]{1988ApJ...333..646E}.
Since the bin size of the gamma-ray data were was set to 7 days,
we also averaged the optical fluxes to match the gamma-ray bins
for the DCF analysis (Section 3.3) and the derivation of ratio
between gamma-ray flux and optical flux.
The unaveraged data were used in all other cases, and specifically in
reporting the polarization degree and angle.
Figure \ref{fig:Sc_3c454} shows
a scatter plot of gamma-ray flux vs optical fluxes for 3C~454.3.  In
this case, this source shows significant correlation between gamma-ray
flux and optical flux with no significant time lag.  We measured the
correlations corresponding to zero time lag for all our samples.  The
error of DCF values are estimated from the variance of the data for
each time-lag interval.  In fact, there are several sources which show
good correlation between gamma-ray and optical fluxes, which actually 
do show time lags 
\citep[e.g., PKS~1510-089][]{2012ApJ...760...69N}.  For PG~1553+113,
correlation between gamma-ray flux and optical flux with time lag is
due to a possible 2-year periodic modulation
\citep{2015ApJ...813L..41A}.  Figure \ref{fig:dcf} shows the DCF plot
for 11 blazars which possess enough data to calculate the correlation
coefficient in several time intervals and show significant
correlations or anti-correlations within the -200 to 200 day time lag
window.  Summary of time lag and correlation coefficient for those 11
blazars are listed in Table \ref{tab:dcf}.
The significance of the correlation (95\% C.L.) was tested using the block Bootstrap method.
In this method, we randomly resampled the data and calculated the correlation coefficient with replacement.
We repeat this routine 10,000 times to get the Bootstrap distribution for each dataset.
From this Bootstrap distribution, we derived a confidence intervals of $\alpha$ = 0.95 (see Appendix B).
Uncertainties in the time lag are derived by the
period that shows significance of correlation (95\% C.L.).
Some of them show asymmetrical shapes in the DCF plot, which might be related
to the difference of rise/decay of the gamma-ray flare and the optical
flare (e.g. 3C~454.3), however, the physical origin of time lags
between gamma-ray and optical features is still unclear
\citep[e.g.,][]{2012ApJ...760..129J} and both delayed and precursory
gamma-ray flares against optical flares are observed in blazars.  To
simplify the discussion and to increase the sample, we use the
correlations corresponding to zero time lag.  We adopt the DCF value
for zero time lag as the correlation index.  We systematically
investigated the correlations between gamma-ray flux and optical flux,
polarization degree and polarized flux with zero time lag.

\begin{table}
  \centering
  \caption{Summary of Correlation time lags between gamma-ray flux and optical flux}
  \label{tab:dcf} 
  \begin{tabular}{lcc}\hline\hline
    Source Name     & time lag  (days)  & DCF peak value  \\ \hline
    AO 0235+164     & 0 $^{+4}_{-14}$     & 0.67 $\pm$ 0.08    \\
    S5 0716+714     & 0 $\pm7$          & 0.47 $\pm$ 0.05    \\
    OJ 287          & -134 $^{+4}_{-28}$      & 1.0  $\pm$ 0.5     \\
    3C 273          & -145 $^{+7}_{-21}$ & -0.97 $\pm$ 0.18   \\
    3C 279          & -28 $\pm 14$   &  0.67 $\pm$ 0.15   \\
    3C 279          & 77 $^{+7}_{-14}$    &  -0.6 $\pm$ 0.1   \\
    PG 1553+113     & 21 $^{+14}_{-28}$  & 0.4 $\pm$ 0.1   \\
    PKS 2155-304    & -28 $^{+28}_{-7}$  & 0.9 $\pm$ 0.2  \\
    BL Lac          & $0^{+28}_{-77}$& 1.0 $\pm$ 0.1   \\
    CTA 102         & 0 $\pm 7$      & 0.8 $\pm$ 0.2 \\
    3C 454.3        & $0^{+49}_{-49}$  & 0.84 $\pm$ 0.13   \\
    \hline
  \end{tabular}
\end{table}

\begin{table*}
  \begin{center}
    \caption{Summary of Correlations}
    \label{tab:res1} 
    \begin{tabular}{lccccccccc}
      \hline\hline
      Source Name    & $L_{\rm gamma}^1$ & $L_{\rm opt}^2$ & $R_{gamma-opt}^3$    & PD$_{\rm max}^4$     & V$_{\rm gamma}^5$   & V$_{\rm opt}^6$  & V$_{\rm PD}^7$ & DCF$_{\rm gamma-opt}^8$ & DCF$_{\rm opt-PD}^9$ \\ \hline
      S2 0109+22     &  45.31         &  45.53        &  0.61 $\pm$ 0.30   &  22.75 $\pm$ 0.69 &  0.39 $\pm$ 0.10  & 0.22 $\pm$ 0.01  & $0.52 \pm 0.02$   & 0.02 $\pm$ 0.09  & -0.25 $\pm$ 0.09 \\
      Mis V1436 &  46.87 &  45.98 &  6.73 $\pm$ 2.45 &  33.68 $\pm$ 1.07 &  0.28 $\pm$ 0.09 &  0.47 $\pm$ 0.01 & $0.52 \pm 0.03$                          & 0.47 $\pm$ 0.23$^*$  &  0.58 $\pm$ 0.07$^*$ \\
      3C 66A &  45.84 &  45.94 &  0.78 $\pm$ 0.28 &  24.76 $\pm$ 0.27 &  0.26 $\pm$ 0.03 &  0.31 $\pm$ 0.01 & $0.39 \pm 0.01$                             & 0.36 $\pm$ 0.09$^*$  &  0.05 $\pm$ 0.02 \\
      AO 0235+164 &  47.40 &  46.58 &  8.16 $\pm$ 1.57 &  33.80 $\pm$ 1.86 &  0.46 $\pm$ 0.03 &  0.83 $\pm$ 0.01 & $0.57 \pm 0.03$                        & 0.67 $\pm$ 0.08$^*$  &  0.84 $\pm$ 0.08$^*$ \\
      PKS 0454-234 &  47.50 &  46.65 &  7.51 $\pm$ 1.18 &  29.98 $\pm$ 1.58 &  0.43 $\pm$ 0.04 &  -- &  $0.61 \pm 0.02$                                   & 0.12 $\pm$ 0.04$^*$  &  0.06 $\pm$ 0.03 \\
      S5 0716+714 &  45.90 &  46.23 &  0.47 $\pm$ 0.13 &  28.53 $\pm$ 0.49 &  0.46 $\pm$ 0.02 &  0.46 $\pm$ 0.01 & $0.52 \pm 0.00$                        & 0.47 $\pm$ 0.05$^*$  & -0.06 $\pm$ 0.02 \\
      OJ 49 &  45.18 &  44.96 &  1.89 $\pm$ 0.80 &  15.76 $\pm$ 0.61 &  0.08 $\pm$ 0.11 &  0.28 $\pm$ 0.01 &  $0.45 \pm 0.04$                             & 0.14 $\pm$ 0.21$^*$  & -0.33 $\pm$ 0.15 \\
      OJ 287 &  45.86 &  45.79 &  1.20 $\pm$ 0.49 &  34.16 $\pm$ 0.38 &  0.54 $\pm$ 0.04 &  0.36 $\pm$ 0.01 & $0.46 \pm 0.01$                             &-0.08 $\pm$ 0.09  &  0.24 $\pm$ 0.05$^*$ \\
      Mrk 421 &  43.62 &  44.43 &  0.16 $\pm$ 0.05 &  5.51 $\pm$ 0.86 &  0.25 $\pm$ 0.04 &  0.30 $\pm$ 0.01 &  $0.50 \pm 0.03$                            &-0.17 $\pm$ 0.41  & -0.39 $\pm$ 0.08 \\
      ON 325 &  44.69 &  44.76 &  0.82 $\pm$ 0.45 &  14.89 $\pm$ 0.25 &  -- & 0.12 $\pm$ 0.01 & $0.29 \pm 0.02$                                           & 0.33 $\pm$ 0.30$^*$  &  0.12 $\pm$ 0.11 \\
      3C 273 &  45.84 &  45.95 &  0.78 $\pm$ 0.18 &  2.37 $\pm$ 0.68 &  0.63 $\pm$ 0.02 &  0.06 $\pm$ 0.01 &  --                                          & 0.13 $\pm$ 0.07  &  0.46 $\pm$ 0.10$^*$ \\
      3C 279 &  46.90 &  45.89 &  11.47 $\pm$ 1.91 &  36.13 $\pm$ 0.10 &  0.88 $\pm$ 0.02 &  0.53 $\pm$ 0.01 &   $0.49 \pm 0.01$                          & 0.17 $\pm$ 0.07  & -0.05 $\pm$ 0.04 \\
      PKS 1502+106 &  48.48 &  46.92 &  39.74 $\pm$ 6.31 &  45.05 $\pm$ 7.24 &  0.43 $\pm$ 0.02 &  0.46 $\pm$ 0.01 &  $0.45 \pm 0.02$                     & 0.66 $\pm$ 0.07$^*$  &  0.49 $\pm$ 0.04$^*$ \\
      PKS 1510-089 &  46.91 &  45.38 &  30.29 $\pm$ 4.16 &  36.13 $\pm$ 1.10 &  0.73 $\pm$ 0.01 &  1.11 $\pm$ 0.01 &  $0.84 \pm 0.04$                     & 0.22 $\pm$ 0.07$^*$  &  0.83 $\pm$ 0.26$^*$ \\
      RX J1542.8+612 &  -- &  -- &  0.60 $\pm$ 0.32 &  15.29 $\pm$ 3.95 &  0.30 $\pm$ 0.09 &  0.16 $\pm$ 0.01 & $0.52 \pm 0.03$                           &-0.37 $\pm$ 0.20$^*$  &  0.09 $\pm$ 0.10 \\
      PG 1553+113 &  45.46 &  46.13 &  0.21 $\pm$ 0.11 &  8.36 $\pm$ 0.38 &  0.62 $\pm$ 0.06 &  0.20 $\pm$ 0.01 &  $0.49 \pm 0.02$                        & 0.10 $\pm$ 0.06$^*$  &  0.03 $\pm$ 0.05 \\
      Mrk 501 &  43.30 &  44.12 &  0.14 $\pm$ 0.09 &  6.45 $\pm$ 1.35 &  0.27 $\pm$ 0.06 &  0.11 $\pm$ 0.01 &  $0.31 \pm 0.03$                            & 0.06 $\pm$ 0.08  &  0.30 $\pm$ 0.11$^*$ \\
      PKS 1749+096 &  45.96 &  45.61 &  2.03 $\pm$ 0.71 &  25.54 $\pm$ 1.79 &  0.24 $\pm$ 0.09 &  0.54 $\pm$ 0.01 &  $0.69 \pm 0.03$                      & 0.89 $\pm$ 0.23  & -0.02 $\pm$ 0.08 \\
      3C 371 &  43.76 &  44.18 &  0.40 $\pm$ 0.21 &  8.80 $\pm$ 1.94 &  -- & 0.18 $\pm$ 0.01 &  $0.29 \pm 0.04$                                           & 0.05 $\pm$ 0.35  & -0.47 $\pm$ 0.33 \\
      1ES 1959+650 &  43.19 &  44.22 &  0.08 $\pm$ 0.07 &  11.39 $\pm$ 1.69 &  0.41 $\pm$ 0.18 &  0.22 $\pm$ 0.01 & $0.37 \pm 0.03$                       & 0.00 $\pm$ 0.09  &  0.02 $\pm$ 0.09 \\
      PKS 2155-304 &  44.77 &  45.38 &  0.22 $\pm$ 0.08 &  8.55 $\pm$ 1.34 &  0.30 $\pm$ 0.04 &  0.31 $\pm$ 0.01 & $0.43 \pm 0.02$                        & 0.41 $\pm$ 0.16$^*$  &  0.49 $\pm$ 0.05$^*$ \\
      BL Lac &  44.72 &  44.83 &  0.77 $\pm$ 0.22 &  25.91 $\pm$ 1.50 &  0.49 $\pm$ 0.02 &  0.44 $\pm$ 0.01 &  $ 0.47 \pm 0.01$                           & 1.02 $\pm$ 0.16$^*$  &  -0.01 $\pm$ 0.04 \\
      CTA 102 &  47.75 &  46.47 &  20.90 $\pm$ 4.70 &  26.93 $\pm$ 0.62 &  0.71 $\pm$ 0.02 &  0.82 $\pm$ 0.01 &  $0.70 \pm 0.03$                          & 0.84 $\pm$ 0.26$^*$ &  0.30 $\pm$ 0.06 \\
      3C 454.3 &  47.92 &  46.60 &  24.32 $\pm$ 2.60 &  33.74 $\pm$ 0.21 &  1.48 $\pm$ 0.01 &  0.74 $\pm$ 0.01 &  $0.79 \pm 0.01$                         & 0.84 $\pm$ 0.13$^*$ &  0.73 $\pm$ 0.04$^*$ \\
      \hline
    \end{tabular}
        {\footnotesize 
          (1-2): Median values of log (luminosity [erg/s]) for gamma-ray and optical luminosity, 
          (3): Median values of ratio between gamma-ray flux and optical flux, errors are derived from 1-$\sigma$ variation of  $R_{gamma-opt}$,
          (4): Maximum polarization in the optical band [\%],
          (5-7): Variability index for gamma-ray, optical band flux and optical polarization degree,
          (8-9): DCF values at timelag = 0 between gamma-ray flux and optical flux, optical flux and polarization degree,
          $^*$: Significant (95\% C.L.) correlation tested using the Bootstrap method.}
  \end{center}
\end{table*}

\subsection{Distribution of gamma-ray luminosity,
optical luminosity and ratio of gamma-ray to optical fluxes}

\begin{figure}
  \centering
  \includegraphics[angle=0,width=9cm]{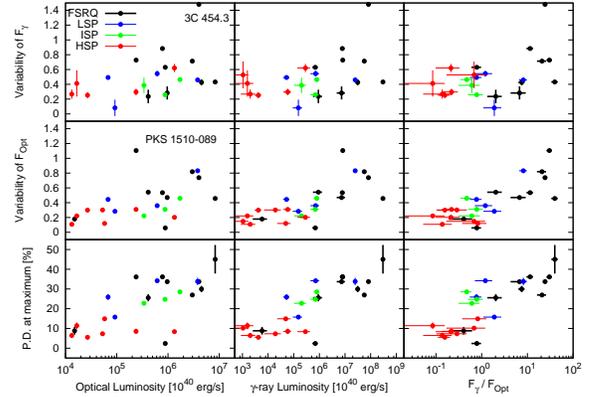}
  \caption{
    Relations between measured properties. From the top to bottom,
    the distribution of variability of gamma-ray flux,
    that of optical flux, and maximum polarization are plotted with
    respect to optical luminosity, gamma-ray luminosity,
    and the ratio of latter to the former are plotted (from left to right).
    The black, blue, green, and red
    symbols indicate FSRQs, LSPs, ISPs, and HSPs, respectively.}
  \label{fig:Sc_Lum_vs_Var}
\end{figure}

Figure \ref{fig:Sc_Lum_vs_Var} shows the distribution of variability indices of 
luminosity and polarization degree calculated separately for each source.
For the luminosity variability index, we calculated the normalized 
``excess variance'' $\sigma^2_{\rm rms}$, \citep[see][]{1997ApJ...476...70N}, described as below,
\begin{equation}
\sigma^2_{\rm rms} = \frac{1}{N\mu^2}\sum_{i=1}^{N}[(X_i-\mu)^2-\sigma_i^2],
\end{equation}
where $N$ is number of observations, $X_i$ and $\sigma_i$ are the data points and their errors, 
and $\mu$ is mean value of $X_i$.
We calculated this for the gamma-ray and optical light curves.
For the polarization degree variability index, we use the maximum observed polarization degree in
Figure \ref{fig:Sc_Lum_vs_Var}.
This is in contrast to Paper I, 
where we used $\Delta{\rm PD} = \max({\rm PD}) - \min({\rm PD})$ as the variability 
index of polarization degree, since most blazars show a minimum polarization degree $\sim 0$\% in our sample.
The highest gamma-ray variability index source is 3C~454.3 (Figure \ref{fig:Sc_Lum_vs_Var}, top line),
and the highest optical variability index source is PKS~1510-089 (Figure \ref{fig:Sc_Lum_vs_Var}, second line).

There is no clear correlation between the variability of gamma-ray flux and the ratio 
of gamma-ray and optical luminosities 
(correlation coefficient of 0.07, see Figure  \ref{fig:Sc_Lum_vs_Var}, right top panel). 
Variability of optical flux and the optical luminosity also does not show clear correlation
(correlation coefficient of 0.42, Figure  \ref{fig:Sc_Lum_vs_Var}, left middle panel).
On the other hand, the variability of optical flux and maximum polarization degree show 
correlations with gamma-ray luminosity and ratio of gamma-ray and optical luminosities 
(correlation coefficient of 0.62$\sim$0.68, Figure  \ref{fig:Sc_Lum_vs_Var}, right bottom panels).
These results imply that the optical luminosity does not play an important role in blazar 
classification.
We note that we did not apply any subtraction of host galaxy component for the optical data.
The contamination of host galaxy changes by the seeing size (equal to aperture size) 
of photometry in the optical band,
and the seeing size at our observatory changes from 1'' to 4'' throughout the year.
It causes an uncertainty in subtracting the host galaxy flux
\citep[typically 10-20\% error, see][]{1999PASP..111.1223N}.
In order to simplify this situation, we did not subtract the host galaxy.
Among these parameters, the ratio of gamma-ray and optical luminosity is a good indicator of 
Compton dominance, as described in the previous section.
What is important in these results is that these correlations
might originate not from variations in the optical luminosity but rather from 
variations in the gamma-ray luminosity.

\begin{figure}
  \centering
  \includegraphics[angle=0,width=9cm]{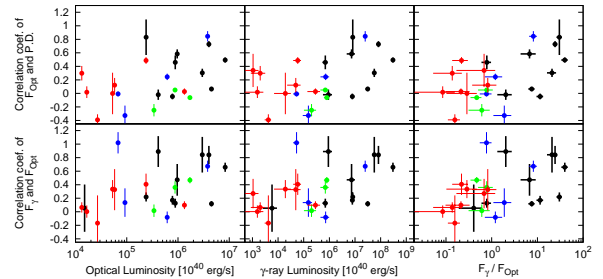}
  \caption{Scatter plot of correlation coefficient and gamma-ray luminosity, 
    optical luminosity and ratio of gamma-ray flux and optical flux. 
    Top panels show DCF value corresponding to zero time lag between optical flux and optical polarization degree.
    Bottom panels show DCF value corresponding to zero time lag between gamma-ray flux and optical flux.
    Colors are the same as those of Figure \ref{fig:Sc_Lum_vs_Var}.}
  \label{fig:Sc_Lum_vs_DCF}
\end{figure}

Figure \ref{fig:Sc_Lum_vs_DCF} shows the distribution of correlation coefficients between gamma-ray and optical luminosities, and between the optical luminosity and optical polarization degree.
We used the DCF value with no time-lag for correlation coefficient (see Section \ref{sec:DCF}).
We find a weak correlation between the DCF value and the gamma-ray luminosity, similar to the case of gamma-ray luminosity vs. maximum polarization degree.
The distribution of correlation between optical flux and optical polarization is 
similar to that reported in Paper I.

\section{Discussion}

\subsection{Summary of our observations}

We collected a large amount of simultaneous gamma-ray and optical 
photopolarimetric data on the variability of blazars.
We confirmed that basic properties, such as a relation between the synchrotron peak frequency
and the amplitude of flux variability, are the same as reported in Paper I.
Paper I suggested that blazars with the peak of the synchrotron radiation located at
higher frequencies had smaller amplitude variations in the flux, color, and polarization degree. 
In addition, we found that some blazars show a significant correlation between the gamma-ray and
optical fluxes, as well as between the optical flux and polarization degree 
\citep[as seen in][]{2009AAS...21441711B,2012ApJ...756...13B}.
In the case of correlation between gamma-ray and optical fluxes,
about 15 out of 24 ($\sim$ 63\%) objects show a significant correlation.
In particular, 7 out of 11 FSRQ blazars
and 8 out of 13 BL~Lac objects show a strong correlation.
This result is consistent with that reported in \cite{2014MNRAS.439..690H}.
We also found a good relation between the correlation coefficient between 
gamma-ray and optical fluxes and the gamma-ray luminosity,
as shown in Fig. \ref{fig:Sc_Lum_vs_DCF}.
A similar relation was found for the correlation coefficient between 
the optical flux and optical polarization degree.  

\subsection{Systematic variation of the maximum polarization degree across blazar sequence}

In this section, we discuss possible origin of the systematic trend in the maximum optical 
polarization degree that is increasing from the HSP ($\sim 10\%$) to the LSP ($\sim 40\%$) blazars.
The maximum optical polarization degree as determined for individual sources appears to be 
fundamentally correlated with either the gamma-ray luminosity or with the Compton dominance 
(here represented by the ratio of gamma-ray to optical fluxes).

The optical emission of most blazars is dominated by synchrotron emission, but in 
some FSRQs in their low state can be contaminated by the thermal emission from the 
accretion disk (in our sample, this seems to be the case for 3C~273).
The synchrotron emission of blazars observed in the optical band is optically thin.
The linear polarization of the optically thin synchrotron radiation depends primarily 
on the structure of magnetic fields in the emitting region, and partially on the energy 
distribution of emitting electrons.

The polarization degree of synchrotron radiation is maximized for uniform magnetic 
fields, and it depends on the electron energy distribution index $p$ 
(such that $N(\gamma) \propto \gamma^{-p}$): $\Pi_{\rm max} = (p+1)/(p+7/3)$ \citep{1959ApJ...130..241W}. 
However, since $\Pi_{\rm max}$ varies between 60\% for $p$ = 1 and 80\% for $p$ = 4,
it is not possible to explain large systematic variations in the polarization degree 
solely by varying the electron distribution function.

Therefore, we need to consider scenarios in which the magnetic fields in the emitting 
regions of FSRQs are systematically better organized than in the case of BL Lacs.
Magnetic fields can be expected to be well organized at the base of relativistic jets, 
where the magnetization parameter ($\sigma_B = \frac{B^2}{4 \pi w}$, where $w$ is the 
specific enthalpy) is well above unity.  
As the jets evolve with distance, their magnetic energy is converted to kinetic energy, 
and they are thought to roughly approach equipartition.
In this condition, it is likely that magnetic fields become tangled by turbulent 
plasma motions, e.g., triggered by current-driven instabilities \citep{1998ApJ...493..291B}.
If the chaotic magnetic field is completely isotropic, it will produce no net synchrotron polarization.
However, as noted by \cite{1980MNRAS.193..439L}, such chaotic fields can be compressed 
by shock waves, resulting in the polarization degree \citep{1985ApJ...298..301H}:
\begin{equation}
\Pi_{\rm max} = \frac{(p+1)}{(p+7/3)}\;\frac{(1-k^2)\cos^2\Phi'}{[2-(1-k^2)\cos^2\Phi']},
\label{eq:pol-degree-shock}
\end{equation}
where $k$ is the shock compression ratio and $\Phi'$ is the inclination of the observer to 
the shock compression plane in the downstream jet co-moving frame.
For example, assuming $p = 2$ and $\Phi' = 0$, the typical maximum polarization degree value 
for FSRQs ($\Pi_{\rm max} \simeq 0.4$) corresponds to $k \simeq 0.5$, and that for 
BL Lacs ($\Pi_{\rm max} \simeq 0.1$) corresponds to $k \simeq 0.9$.
This would suggest very weak shock waves in the case of BL Lacs,
potentially creating a problem for efficient particle acceleration.
The distribution of viewing angles in the co-moving frame can
be expected to be roughly isotropic, hence, it is very unlikely
that the $\Pi_{\rm max}$ values could be reduced at low shock
compression ratios due to a specific choice of $\Phi'$ values.
If there would be strong shock waves with very low compression ratios,
we should observe even higher polarization degrees in some blazars.
Therefore, we think that variations in the shock compression ratio
cannot reasonably explain the differences in maximum polarization degree 
across different types of blazars.

Depolarization of synchrotron radiation could result from a superposition 
of multiple emitting regions with independent orientations of magnetic field lines \citep{1985ApJ...290..627J}.
In such a case, BL Lacs should be characterized by a larger number of emitting regions.
This would also predict a smaller variability index.
In fact, our results indicate the optical variability index is correlated with the 
Compton dominance, but no such trend is apparent for the gamma-ray
variability index (see Figure \ref{fig:Sc_Lum_vs_Var}).

Such multiple emitting regions in blazars might be characterized by the distribution of 
electron energies varying from one region to another.  The typical electron energy 
$\gamma_{\rm opt} \propto [(1+z)\nu_{\rm opt}/(\Gamma_{\rm j}B')]^{1/2}$ corresponding to the 
optical band is somewhat higher in the case of BL Lacs than in FSRQs.
This is because FSRQs show higher Lorentz factors $\Gamma_{\rm j}$ \citep{2009A&A...494..527H} and 
stronger magnetic fields $B'$ \citep{2012A&A...545A.113P} than BL Lacs, although these 
differences are not large.  However, the electron energy distribution extends to a 
significantly higher maximum characteristic energy $\gamma_{\rm max} \gg \gamma_{\rm opt}$ 
in the case of BL Lacs (where the synchrotron component extends to the X-ray band) than 
in the case of FSRQs (where $\gamma_{\rm max} \sim \gamma_{\rm opt}$).
If the number of emitting regions or the volume filling factor scale 
with $\gamma/\gamma_{\rm max}$, this could explain a lower effective polarization degree of
BL Lacs \cite[e.g.,][]{2010arXiv1005.5551M}.
This predicts that the variability index scales with $\gamma/\gamma_{\rm max}$ , 
and indeed there is some observational evidence that this 
is the case \citep[e.g.,][]{2015A&A...578A..22A}.
This also predicts a high X-ray polarization degree for BL~Lac objects, 
which can be verified by future X-ray polarimetric missions.

A particular scenario that could explain our main result is a spine-sheath 
model \citep{2005A&A...432..401G}, in which the fast spine has ordered magnetic 
fields and the slow sheath has chaotic magnetic fields.
In this scenario, the spine region would produce a highly variable and polarized 
synchrotron component, and the sheath region would produce a steady and weakly polarized component.
In order to explain the systematic trend in maximum polarization degree, the jet 
volume fraction occupied by the spine should increase from the BL Lacs to the FSRQ blazars.

These scenarios should be related to the fundamental differences between the 
relativistic jets of FSRQs and BL Lacs.  In the Unification Model for AGNs 
\citep{1995PASP..107..803U}, FSRQs are associated with powerful FR II radio galaxies 
\citep{1974MNRAS.167P..31F}, and BL Lacs are associated with relatively weak FR I radio galaxies.
FR II jets are known to form strong hotspots, which indicates that they carry a 
relatively large fraction of their initial kinetic power to distances $>100\;{\rm kpc}$.
On the other hand, FR I jets appear to gradually dissipate their kinetic energy, 
so that they do not form hotspots.  This suggests that FR I jets are more turbulent, 
and therefore their magnetic fields are more chaotic and less organized, than FR II jets \citep[e.g.,][]{2016MNRAS.461L..46T}.
This fundamental dichotomy in the properties of relativistic jets naturally explains 
our observational result that the maximum optical polarization degree is 
systematically higher in the FSRQs as compared to the BL Lacs.
However, most of the blazar emission is expected to be produced roughly on 
pc scales \citep[e.g.,][]{2014ApJ...789..161N}, and hence this dichotomy (FRI or FRII) 
in jet properties should manifest itself already at these scales.

\section{Conclusion}

We performed long-term photopolarimetric monitoring of GeV bright blazars detected 
by the Fermi-LAT using Kanata telescope for 6.5 years.  
We selected 45 blazars of various sub-types, and obtained densely-sampled 
simultaneous light curves in the optical and GeV band for 24 blazars.
Our results are
(1) some blazars show a significant correlation between the gamma-ray and
optical fluxes, as well as between the optical flux and polarization degree,
(2) a significant correlation between the maximum degree of optical 
linear polarization  and the gamma-ray luminosity.  
These relations are also confirmed with the ratio of gamma-ray to optical 
fluxes instead of gamma-ray luminosity.
These results can be explained by a spine-sheath model and 
systematic difference in the intrinsic alignment of magnetic fields in 
relativistic jets (e.g., FSRQs vs. BL Lacs or FR Is vs. FR IIs).
A measurement of flare amplitude and frequency could be related with size and number 
of emission regions in the jet, therefore such  a measurement of ``Flare cadence'' 
will be helpful to test the assumptions of this model.

\section*{Acknowledgement}

The Fermi LAT Collaboration acknowledges generous
ongoing support from a number of agencies and institutes
that have supported both the development and the operation
of the LAT as well as scientific data analysis. These
include the National Aeronautics and Space Administration
and the Department of Energy in the United States, the
Commissariat `a lfEnergie Atomique and the Centre National
de la Recherche Scientifique/Institut National de Physique
NuclLeaire et de Physique des Particules in France, the Agenzia
Spaziale Italiana and the Istituto Nazionale di Fisica Nucleare
in Italy, the Ministry of Education, Culture, Sports, Science
and Technology (MEXT), High Energy Accelerator Research
Organization (KEK) and Japan Aerospace Exploration Agency
(JAXA) in Japan, and the K. A. Wallenberg Foundation, the
Swedish Research Council and the Swedish National Space
Board in Sweden.
This work was supported by JSPS KAKENHI Grant Numbers 24000004, 24244014.
This work was supported by JSPS and NSF under the JSPS-NSF Partnerships
for International Research and Education (PIRE). 
This work was partly supported by the Hirao Taro Foundation
of the Konan University Association for Academic Research.

\bibliographystyle{apj}

\begin{thebibliography}{44}
\expandafter\ifx\csname natexlab\endcsname\relax\def\natexlab#1{#1}\fi

\bibitem[{{Acero} {et~al.}(2015){Acero}, {Ackermann}, {Ajello}, {Albert},
  {Atwood}, {Axelsson}, {Baldini}, {Ballet}, {Barbiellini}, {Bastieri},
  {Belfiore}, {Bellazzini}, {Bissaldi}, {Blandford}, {Bloom}, {Bogart},
  {Bonino}, {Bottacini}, {Bregeon}, {Britto}, {Bruel}, {Buehler}, {Burnett},
  {Buson}, {Caliandro}, {Cameron}, {Caputo}, {Caragiulo}, {Caraveo},
  {Casandjian}, {Cavazzuti}, {Charles}, {Chaves}, {Chekhtman}, {Cheung},
  {Chiang}, {Chiaro}, {Ciprini}, {Claus}, {Cohen-Tanugi}, {Cominsky}, {Conrad},
  {Cutini}, {DAmmando}, {de Angelis}, {DeKlotz}, {de Palma}, {Desiante},
  {Digel}, {Di Venere}, {Drell}, {Dubois}, {Dumora}, {Favuzzi}, {Fegan},
  {Ferrara}, {Finke}, {Franckowiak}, {Fukazawa}, {Funk}, {Fusco}, {Gargano},
  {Gasparrini}, {Giebels}, {Giglietto}, {Giommi}, {Giordano}, {Giroletti},
  {Glanzman}, {Godfrey}, {Grenier}, {Grondin}, {Grove}, {Guillemot}, {Guiriec},
  {Hadasch}, {Harding}, {Hays}, {Hewitt}, {Hill}, {Horan}, {Iafrate}, {Jogler},
  {J{\'o}hannesson}, {Johnson}, {Johnson}, {Johnson}, {Johnson}, {Kamae},
  {Kataoka}, {Katsuta}, {Kuss}, {La Mura}, {Landriu}, {Larsson}, {Latronico},
  {Lemoine-Goumard}, {Li}, {Li}, {Longo}, {Loparco}, {Lott}, {Lovellette},
  {Lubrano}, {Madejski}, {Massaro}, {Mayer}, {Mazziotta}, {McEnery},
  {Michelson}, {Mirabal}, {Mizuno}, {Moiseev}, {Mongelli}, {Monzani},
  {Morselli}, {Moskalenko}, {Murgia}, {Nuss}, {Ohno}, {Ohsugi}, {Omodei},
  {Orienti}, {Orlando}, {Ormes}, {Paneque}, {Panetta}, {Perkins},
  {Pesce-Rollins}, {Piron}, {Pivato}, {Porter}, {Racusin}, {Rando}, {Razzano},
  {Razzaque}, {Reimer}, {Reimer}, {Reposeur}, {Rochester}, {Romani},
  {Salvetti}, {S{\'a}nchez-Conde}, {Saz Parkinson}, {Schulz}, {Siskind},
  {Smith}, {Spada}, {Spandre}, {Spinelli}, {Stephens}, {Strong}, {Suson},
  {Takahashi}, {Takahashi}, {Tanaka}, {Thayer}, {Thayer}, {Thompson},
  {Tibaldo}, {Tibolla}, {Torres}, {Torresi}, {Tosti}, {Troja}, {Van Klaveren},
  {Vianello}, {Winer}, {Wood}, {Wood}, \& {Zimmer}}]{2015ApJS..218...23A}
  {Acero}, F., {Ackermann}, M., {Ajello}, M., {et~al.} 2015, \apjs, 218, 23

\bibitem[{{Acero} {et~al.}(2016){Acero}, {Ackermann}, {Ajello}, {Albert},
  {Baldini}, {Ballet}, {Barbiellini}, {Bastieri}, {Bellazzini}, {Bissaldi},
  {Bloom}, {Bonino}, {Bottacini}, {Brandt}, {Bregeon}, {Bruel}, {Buehler},
  {Buson}, {Caliandro}, {Cameron}, {Caragiulo}, {Caraveo}, {Casandjian},
  {Cavazzuti}, {Cecchi}, {Charles}, {Chekhtman}, {Chiang}, {Chiaro}, {Ciprini},
  {Claus}, {Cohen-Tanugi}, {Conrad}, {Cuoco}, {Cutini}, {D'Ammando}, {de
  Angelis}, {de Palma}, {Desiante}, {Digel}, {Di Venere}, {Drell}, {Favuzzi},
  {Fegan}, {Ferrara}, {Focke}, {Franckowiak}, {Funk}, {Fusco}, {Gargano},
  {Gasparrini}, {Giglietto}, {Giordano}, {Giroletti}, {Glanzman}, {Godfrey},
  {Grenier}, {Guiriec}, {Hadasch}, {Harding}, {Hayashi}, {Hays}, {Hewitt},
  {Hill}, {Horan}, {Hou}, {Jogler}, {J{\'o}hannesson}, {Kamae}, {Kuss},
  {Landriu}, {Larsson}, {Latronico}, {Li}, {Li}, {Longo}, {Loparco},
  {Lovellette}, {Lubrano}, {Maldera}, {Malyshev}, {Manfreda}, {Martin},
  {Mayer}, {Mazziotta}, {McEnery}, {Michelson}, {Mirabal}, {Mizuno}, {Monzani},
  {Morselli}, {Nuss}, {Ohsugi}, {Omodei}, {Orienti}, {Orlando}, {Ormes},
  {Paneque}, {Pesce-Rollins}, {Piron}, {Pivato}, {Rain{\`o}}, {Rando},
  {Razzano}, {Razzaque}, {Reimer}, {Reimer}, {Remy}, {Renault},
  {S{\'a}nchez-Conde}, {Schaal}, {Schulz}, {Sgr{\`o}}, {Siskind}, {Spada},
  {Spandre}, {Spinelli}, {Strong}, {Suson}, {Tajima}, {Takahashi}, {Thayer},
  {Thompson}, {Tibaldo}, {Tinivella}, {Torres}, {Tosti}, {Troja}, {Vianello},
  {Werner}, {Wood}, {Wood}, {Zaharijas}, \& {Zimmer}}]{2016ApJS..223...26A}
  {Acero}, F., {Ackermann}, M., {Ajello}, M., {et~al.} 2016, \apjs, 223, 26

\bibitem[{{Ackermann} {et~al.}(2011){Ackermann}, {Ajello}, {Allafort},
  {Antolini}, {Atwood}, {Axelsson}, {Baldini}, {Ballet}, {Barbiellini},
  {Bastieri}, {Bechtol}, {Bellazzini}, {Berenji}, {Blandford}, {Bloom},
  {Bonamente}, {Borgland}, {Bottacini}, {Bouvier}, {Bregeon}, {Brigida},
  {Bruel}, {Buehler}, {Burnett}, {Buson}, {Caliandro}, {Cameron}, {Caraveo},
  {Casandjian}, {Cavazzuti}, {Cecchi}, {Charles}, {Cheung}, {Chiang},
  {Ciprini}, {Claus}, {Cohen-Tanugi}, {Conrad}, {Costamante}, {Cutini}, {de
  Angelis}, {de Palma}, {Dermer}, {Digel}, {Silva}, {Drell}, {Dubois},
  {Escande}, {Favuzzi}, {Fegan}, {Ferrara}, {Finke}, {Focke}, {Fortin},
  {Frailis}, {Fukazawa}, {Funk}, {Fusco}, {Gargano}, {Gasparrini}, {Gehrels},
  {Germani}, {Giebels}, {Giglietto}, {Giommi}, {Giordano}, {Giroletti},
  {Glanzman}, {Godfrey}, {Grenier}, {Grove}, {Guiriec}, {Gustafsson},
  {Hadasch}, {Hayashida}, {Hays}, {Healey}, {Horan}, {Hou}, {Hughes},
  {Iafrate}, {J{\'o}hannesson}, {Johnson}, {Johnson}, {Kamae}, {Katagiri},
  {Kataoka}, {Kn{\"o}dlseder}, {Kuss}, {Lande}, {Larsson}, {Latronico},
  {Longo}, {Loparco}, {Lott}, {Lovellette}, {Lubrano}, {Madejski}, {Mazziotta},
  {McConville}, {McEnery}, {Michelson}, {Mitthumsiri}, {Mizuno}, {Moiseev},
  {Monte}, {Monzani}, {Moretti}, {Morselli}, {Moskalenko}, {Murgia},
  {Nakamori}, {Naumann-Godo}, {Nolan}, {Norris}, {Nuss}, {Ohno}, {Ohsugi},
  {Okumura}, {Omodei}, {Orienti}, {Orlando}, {Ormes}, {Ozaki}, {Paneque},
  {Parent}, {Pesce-Rollins}, {Pierbattista}, {Piranomonte}, {Piron}, {Pivato},
  {Porter}, {Rain{\`o}}, {Rando}, {Razzano}, {Razzaque}, {Reimer}, {Reimer},
  {Ritz}, {Rochester}, {Romani}, {Roth}, {Sanchez}, {Sbarra}, {Scargle},
  {Schalk}, {Sgr{\`o}}, {Shaw}, {Siskind}, {Spandre}, {Spinelli}, {Strong},
  {Suson}, {Tajima}, {Takahashi}, {Takahashi}, {Tanaka}, {Thayer}, {Thayer},
  {Thompson}, {Tibaldo}, {Tinivella}, {Torres}, {Tosti}, {Troja}, \&
  {Zimmer}}]{2011ApJ...743..171A}
  {Ackermann}, M., {Ajello}, M., {Allafort}, A., {et~al.} 2011, \apj, 743, 171

\bibitem[{{Ackermann} {et~al.}(2015{\natexlab{a}}){Ackermann}, {Ajello},
  {Albert}, {Atwood}, {Baldini}, {Ballet}, {Barbiellini}, {Bastieri}, {Becerra
  Gonzalez}, {Bellazzini}, {Bissaldi}, {Blandford}, {Bloom}, {Bonino},
  {Bottacini}, {Bregeon}, {Bruel}, {Buehler}, {Buson}, {Caliandro}, {Cameron},
  {Caputo}, {Caragiulo}, {Caraveo}, {Cavazzuti}, {Cecchi}, {Chekhtman},
  {Chiang}, {Chiaro}, {Ciprini}, {Cohen-Tanugi}, {Conrad}, {Cutini},
  {D'Ammando}, {de Angelis}, {de Palma}, {Desiante}, {Di Venere},
  {Domi{\acute}nguez}, {Drell}, {Favuzzi}, {Fegan}, {Ferrara}, {Focke},
  {Fuhrmann}, {Fukazawa}, {Fusco}, {Gargano}, {Gasparrini}, {Giglietto},
  {Giommi}, {Giordano}, {Giroletti}, {Godfrey}, {Green}, {Grenier}, {Grove},
  {Guiriec}, {Harding}, {Hays}, {Hewitt}, {Hill}, {Horan}, {Jogler},
  {J{\'o}hannesson}, {Johnson}, {Kamae}, {Kuss}, {Larsson}, {Latronico}, {Li},
  {Li}, {Longo}, {Loparco}, {Lott}, {Lovellette}, {Lubrano}, {Magill},
  {Maldera}, {Manfreda}, {Max-Moerbeck}, {Mayer}, {Mazziotta}, {McEnery},
  {Michelson}, {Mizuno}, {Monzani}, {Morselli}, {Moskalenko}, {Murgia}, {Nuss},
  {Ohno}, {Ohsugi}, {Ojha}, {Omodei}, {Orlando}, {Ormes}, {Paneque}, {Pearson},
  {Perkins}, {Perri}, {Pesce-Rollins}, {Petrosian}, {Piron}, {Pivato},
  {Porter}, {Rain{\`o}}, {Rando}, {Razzano}, {Readhead}, {Reimer}, {Reimer},
  {Schulz}, {Sgr{\`o}}, {Siskind}, {Spada}, {Spandre}, {Spinelli}, {Suson},
  {Takahashi}, {Thayer}, {Thompson}, {Tibaldo}, {Torres}, {Tosti}, {Troja},
  {Uchiyama}, {Vianello}, {Wood}, {Wood}, {Zimmer}, {Berdyugin}, {Corbet},
  {Hovatta}, {Lindfors}, {Nilsson}, {Reinthal}, {Sillanp{\"a}{\"a}},
  {Stamerra}, {Takalo}, \& {Valtonen}}]{2015ApJ...813L..41A}
  {Ackermann}, M., {Ajello}, M., {Albert}, A., {et~al.} 2015{\natexlab{a}}, \apjl, 813, L41

\bibitem[{{Ackermann} {et~al.}(2015{\natexlab{b}}){Ackermann}, {Ajello},
  {Atwood}, {Baldini}, {Ballet}, {Barbiellini}, {Bastieri}, {Becerra Gonzalez},
  {Bellazzini}, {Bissaldi}, {Blandford}, {Bloom}, {Bonino}, {Bottacini},
  {Brandt}, {Bregeon}, {Britto}, {Bruel}, {Buehler}, {Buson}, {Caliandro},
  {Cameron}, {Caragiulo}, {Caraveo}, {Carpenter}, {Casandjian}, {Cavazzuti},
  {Cecchi}, {Charles}, {Chekhtman}, {Cheung}, {Chiang}, {Chiaro}, {Ciprini},
  {Claus}, {Cohen-Tanugi}, {Cominsky}, {Conrad}, {Cutini}, {D'Abrusco},
  {D'Ammando}, {de Angelis}, {Desiante}, {Digel}, {Di Venere}, {Drell},
  {Favuzzi}, {Fegan}, {Ferrara}, {Finke}, {Focke}, {Franckowiak}, {Fuhrmann},
  {Fukazawa}, {Furniss}, {Fusco}, {Gargano}, {Gasparrini}, {Giglietto},
  {Giommi}, {Giordano}, {Giroletti}, {Glanzman}, {Godfrey}, {Grenier}, {Grove},
  {Guiriec}, {Hewitt}, {Hill}, {Horan}, {Itoh}, {J{\'o}hannesson}, {Johnson},
  {Johnson}, {Kataoka}, {Kawano}, {Krauss}, {Kuss}, {La Mura}, {Larsson},
  {Latronico}, {Leto}, {Li}, {Li}, {Longo}, {Loparco}, {Lott}, {Lovellette},
  {Lubrano}, {Madejski}, {Mayer}, {Mazziotta}, {McEnery}, {Michelson},
  {Mizuno}, {Moiseev}, {Monzani}, {Morselli}, {Moskalenko}, {Murgia}, {Nuss},
  {Ohno}, {Ohsugi}, {Ojha}, {Omodei}, {Orienti}, {Orlando}, {Paggi}, {Paneque},
  {Perkins}, {Pesce-Rollins}, {Piron}, {Pivato}, {Porter}, {Rain{\`o}},
  {Rando}, {Razzano}, {Razzaque}, {Reimer}, {Reimer}, {Romani}, {Salvetti},
  {Schaal}, {Schinzel}, {Schulz}, {Sgr{\`o}}, {Siskind}, {Sokolovsky}, {Spada},
  {Spandre}, {Spinelli}, {Stawarz}, {Suson}, {Takahashi}, {Takahashi},
  {Tanaka}, {Thayer}, {Thayer}, {Tibaldo}, {Torres}, {Torresi}, {Tosti},
  {Troja}, {Uchiyama}, {Vianello}, {Winer}, {Wood}, \&
  {Zimmer}}]{2015ApJ...810...14A}
  {Ackermann}, M., {Ajello}, M., {Atwood}, W.~B., {et~al.} 2015{\natexlab{b}}, \apj, 810, 14

\bibitem[{{Agudo} {et~al.}(2011){Agudo}, {Marscher}, {Jorstad}, {Larionov},
  {G{\'o}mez}, {L{\"a}hteenm{\"a}ki}, {Smith}, {Nilsson}, {Readhead}, {Aller},
  {Heidt}, {Gurwell}, {Thum}, {Wehrle}, {Nikolashvili}, {Aller},
  {Ben{\'{\i}}tez}, {Blinov}, {Hagen-Thorn}, {Hiriart}, {Jannuzi}, {Joshi},
  {Kimeridze}, {Kurtanidze}, {Kurtanidze}, {Lindfors}, {Molina}, {Morozova},
  {Nieppola}, {Olmstead}, {Reinthal}, {Roca-Sogorb}, {Schmidt}, {Sigua},
  {Sillanp{\"a}{\"a}}, {Takalo}, {Taylor}, {Tornikoski}, {Troitsky}, {Zook}, \&
  {Wiesemeyer}}]{2011ApJ...735L..10A}
  {Agudo}, I., {Marscher}, A.~P., {Jorstad}, S.~G., {et~al.} 2011, \apjl, 735, L10

\bibitem[{{Aleksi{\'c}} {et~al.}(2015){Aleksi{\'c}}, {Ansoldi}, {Antonelli},
  {Antoranz}, {Babic}, {Bangale}, {Barres de Almeida}, {Barrio}, {Becerra
  Gonz{\'a}lez}, {Bednarek}, \& et~al.}]{2015A&A...578A..22A}
{Aleksi{\'c}}, J., {Ansoldi}, S., {Antonelli}, L.~A., {et~al.} 2015, \aap, 578, A22

\bibitem[{{Atwood} {et~al.}(2009){Atwood}, {Abdo}, {Ackermann}, {Althouse},
  {Anderson}, {Axelsson}, {Baldini}, {Ballet}, {Band}, {Barbiellini}, \&
  et~al.}]{2009ApJ...697.1071A}
  {Atwood}, W.~B., {Abdo}, A.~A., {Ackermann}, M., {et~al.} 2009, \apj, 697, 1071

\bibitem[{{Begelman}(1998)}]{1998ApJ...493..291B}
  {Begelman}, M.~C. 1998, \apj, 493, 291

\bibitem[{{Blinov} {et~al.}(2015){Blinov}, {Pavlidou}, {Papadakis},
  {Kiehlmann}, {Panopoulou}, {Liodakis}, {King}, {Angelakis}, {Balokovi{\'c}},
  {Das}, {Feiler}, {Fuhrmann}, {Hovatta}, {Khodade}, {Kus}, {Kylafis},
  {Mahabal}, {Myserlis}, {Modi}, {Pazderska}, {Pazderski}, {Papamastorakis},
  {Pearson}, {Rajarshi}, {Ramaprakash}, {Reig}, {Readhead}, {Tassis}, \&
  {Zensus}}]{2015MNRAS.453.1669B}
  {Blinov}, D., {Pavlidou}, V., {Papadakis}, I., {et~al.} 2015, \mnras, 453, 1669

\bibitem[{{Bonning} {et~al.}(2012){Bonning}, {Urry}, {Bailyn}, {Buxton},
  {Chatterjee}, {Coppi}, {Fossati}, {Isler}, \&
  {Maraschi}}]{2012ApJ...756...13B}
  {Bonning}, E., {Urry}, C.~M., {Bailyn}, C., {et~al.} 2012, \apj, 756, 13

\bibitem[{{Bonning} {et~al.}(2009){Bonning}, {Bailyn}, {Urry}, {Buxton},
  {Fossati}, {Maraschi}, {Coppi}, {Scalzo}, {Isler}, \&
  {Kaptur}}]{2009AAS...21441711B}
  {Bonning}, E.~W., {Bailyn}, C., {Urry}, C., {et~al.} 2009, in American Astronomical Society Meeting
  Abstracts, Vol. 214, American Astronomical Society Meeting Abstracts 214, 686

\bibitem[{{Brand}(1985)}]{1985agn..book..215B}
  {Brand}, P.~W.~J.~L. 1985, {Infrared and optical photopolarimetry of blazars},
  ed. T.~{Neckel} \& H.~{Vehrenberg}, 215--219

\bibitem[{{Edelson} \& {Krolik}(1988)}]{1988ApJ...333..646E}
  {Edelson}, R.~A., \& {Krolik}, J.~H. 1988, \apj, 333, 646

\bibitem[{{Fanaroff} \& {Riley}(1974)}]{1974MNRAS.167P..31F}
  {Fanaroff}, B.~L., \& {Riley}, J.~M. 1974, \mnras, 167, 31P

\bibitem[{{Finke}(2013)}]{2013ApJ...763..134F}
  {Finke}, J.~D. 2013, \apj, 763, 134

\bibitem[{{Fossati} {et~al.}(1998){Fossati}, {Maraschi}, {Celotti}, {Comastri},
  \& {Ghisellini}}]{1998MNRAS.299..433F}
  {Fossati}, G., {Maraschi}, L., {Celotti}, A., {et~al.} 1998, \mnras, 299, 433

\bibitem[{{Ghisellini} {et~al.}(1998){Ghisellini}, {Celotti}, {Fossati},
    {Maraschi}, \& {Comastri}}]{1998MNRAS.301..451G}
  {Ghisellini}, G., {Celotti}, A., {Fossati}, G., {et~al.} 1998, \mnras, 301, 451

\bibitem[{{Ghisellini} {et~al.}(2005){Ghisellini}, {Tavecchio}, \&
    {Chiaberge}}]{2005A&A...432..401G}
  {Ghisellini}, G., {Tavecchio}, F., \& {Chiaberge}, M. 2005, \aap, 432, 401

\bibitem[{{Hovatta} {et~al.}(2009){Hovatta}, {Valtaoja}, {Tornikoski}, \&
    {L{\"a}hteenm{\"a}ki}}]{2009A&A...494..527H}
  {Hovatta}, T., {Valtaoja}, E., {Tornikoski}, M., {et~al.} 2009, \aap, 494, 527

\bibitem[{{Hovatta} {et~al.}(2014){Hovatta}, {Pavlidou}, {King}, {Mahabal},
  {Sesar}, {Dancikova}, {Djorgovski}, {Drake}, {Laher}, {Levitan},
  {Max-Moerbeck}, {Ofek}, {Pearson}, {Prince}, {Readhead}, {Richards}, \&
  {Surace}}]{2014MNRAS.439..690H}
  {Hovatta}, T., {Pavlidou}, V., {King}, O.~G., {et~al.} 2014, \mnras, 439, 690

\bibitem[{{Hughes} {et~al.}(1985){Hughes}, {Aller}, \&
  {Aller}}]{1985ApJ...298..301H}
  {Hughes}, P.~A., {Aller}, H.~D., \& {Aller}, M.~F. 1985, \apj, 298, 301

\bibitem[{{Ikejiri} {et~al.}(2011){Ikejiri}, {Uemura}, {Sasada}, {Ito},
  {Yamanaka}, {Sakimoto}, {Arai}, {Fukazawa}, {Ohsugi}, {Kawabata}, {Yoshida},
  {Sato}, \& {Kino}}]{2011PASJ...63..639I}
  {Ikejiri}, Y., {Uemura}, M., {Sasada}, M., {et~al.} 2011, \pasj, 63, 639

\bibitem[{{Janiak} {et~al.}(2012){Janiak}, {Sikora}, {Nalewajko}, {Moderski},
  \& {Madejski}}]{2012ApJ...760..129J}
  {Janiak}, M., {Sikora}, M., {Nalewajko}, K., {et~al.} 2012, \apj, 760, 129

\bibitem[{Jones} {et~al.}(1985)]{1985ApJ...290..627J}
  {Jones}, T.~W., {Rudnick}, L., {Fiedler}, R.~L., {Aller}, H.~D., {Aller}, M.~F., \& {Hodge}, P.~E., 1985, ApJ, 290, 627

\bibitem[{{Kawabata} {et~al.}(2008){Kawabata}, {Nagae}, {Chiyonobu}, {Tanaka},
  {Nakaya}, {Suzuki}, {Kamata}, {Miyazaki}, {Hiragi}, {Miyamoto}, {Yamanaka},
  {Arai}, {Yamashita}, {Uemura}, {Ohsugi}, {Isogai}, {Ishitobi}, \&
  {Sato}}]{2008SPIE.7014E.151K}
  {Kawabata}, K.~S., {Nagae}, O., {Chiyonobu}, S., {et~al.} 2008, in Society of Photo-Optical Instrumentation
  Engineers (SPIE) Conference Series, Vol. 7014, Society of Photo-Optical
  Instrumentation Engineers (SPIE) Conference Series

\bibitem[{{Kiehlmann} {et~al.}(2016){Kiehlmann}, {Savolainen}, {Jorstad},
  {Sokolovsky}, {Schinzel}, {Marscher}, {Larionov}, {Agudo}, {Akitaya},
  {Ben{\'{\i}}tez}, {Berdyugin}, {Blinov}, {Bochkarev}, {Borman}, {Burenkov},
  {Casadio}, {Doroshenko}, {Efimova}, {Fukazawa}, {G{\'o}mez}, {Grishina},
  {Hagen-Thorn}, {Heidt}, {Hiriart}, {Itoh}, {Joshi}, {Kawabata}, {Kimeridze},
  {Kopatskaya}, {Korobtsev}, {Krajci}, {Kurtanidze}, {Kurtanidze}, {Larionova},
  {Larionova}, {Lindfors}, {L{\'o}pez}, {McHardy}, {Molina}, {Moritani},
  {Morozova}, {Nazarov}, {Nikolashvili}, {Nilsson}, {Pulatova}, {Reinthal},
  {Sadun}, {Sasada}, {Savchenko}, {Sergeev}, {Sigua}, {Smith}, {Sorcia},
  {Spiridonova}, {Takaki}, {Takalo}, {Taylor}, {Troitsky}, {Uemura},
  {Ugolkova}, {Ui}, {Yoshida}, {Zensus}, \& {Zhdanova}}]{2016A&A...590A..10K}
  {Kiehlmann}, S., {Savolainen}, T., {Jorstad}, S.~G., {et~al.} 2016, \aap, 590, A10

\bibitem[{{Laing}(1980)}]{1980MNRAS.193..439L}
  {Laing}, R.~A. 1980, \mnras, 193, 439

\bibitem[{{Loh}(2008)}]{2008ApJ...681..726L}
  {Loh}, J.~M. 2008, \apj, 681, 726

\bibitem[{{Marscher} \& {Jorstad}(2010)}]{2010arXiv1005.5551M}
  {Marscher}, A.~P., \& {Jorstad}, S.~G. 2010, ArXiv e-prints

\bibitem[{{Marscher} {et~al.}(2008){Marscher}, {Jorstad}, {D'Arcangelo},
  {Smith}, {Williams}, {Larionov}, {Oh}, {Olmstead}, {Aller}, {Aller},
  {McHardy}, {L{\"a}hteenm{\"a}ki}, {Tornikoski}, {Valtaoja}, {Hagen-Thorn},
  {Kopatskaya}, {Gear}, {Tosti}, {Kurtanidze}, {Nikolashvili}, {Sigua},
  {Miller}, \& {Ryle}}]{2008Natur.452..966M}
  {Marscher}, A.~P., {Jorstad}, S.~G., {D'Arcangelo}, F.~D., {et~al.} 2008, \nat, 452, 966

\bibitem[{{Mead} {et~al.}(1990){Mead}, {Ballard}, {Brand}, {Hough}, {Brindle},
  \& {Bailey}}]{1990A&AS...83..183M}
{Mead}, A.~R.~G., {Ballard}, K.~R., {Brand}, P.~W.~J.~L., {Hough}, J.~H.,
  {Brindle}, C., \& {Bailey}, J.~A. 1990, \aaps, 83, 183

\bibitem[{{Nalewajko} {et~al.}(2014){Nalewajko}, {Begelman}, \&
  {Sikora}}]{2014ApJ...789..161N}
{Nalewajko}, K., {Begelman}, M.~C., \& {Sikora}, M. 2014, \apj, 789, 161

\bibitem[{{Nalewajko} {et~al.}(2012){Nalewajko}, {Sikora}, {Madejski}, {Exter},
  {Szostek}, {Szczerba}, {Kidger}, \& {Lorente}}]{2012ApJ...760...69N}
{Nalewajko}, K., {Sikora}, M., {Madejski}, G.~M., {Exter}, K., {Szostek}, A.,
  {Szczerba}, R., {Kidger}, M.~R., \& {Lorente}, R. 2012, \apj, 760, 69

\bibitem[{{Nandra} {et~al.}(1997){Nandra}, {George}, {Mushotzky}, {Turner}, \&
  {Yaqoob}}]{1997ApJ...476...70N}
{Nandra}, K., {George}, I.~M., {Mushotzky}, R.~F., {Turner}, T.~J., \&
  {Yaqoob}, T. 1997, \apj, 476, 70


\bibitem[Nilsson et al.(1999)]{1999PASP..111.1223N} 
  Nilsson, K., Pursimo, T., Takalo, L.~O., et al.\ 1999, \pasp, 111, 1223 


\bibitem[{{Pushkarev} {et~al.}(2012){Pushkarev}, {Hovatta}, {Kovalev},
  {Lister}, {Lobanov}, {Savolainen}, \& {Zensus}}]{2012A&A...545A.113P}
{Pushkarev}, A.~B., {Hovatta}, T., {Kovalev}, Y.~Y., {Lister}, M.~L.,
  {Lobanov}, A.~P., {Savolainen}, T., \& {Zensus}, J.~A. 2012, \aap, 545, A113

\bibitem[{{Sasada} {et~al.}(2008){Sasada}, {Uemura}, {Arai}, {Fukazawa},
  {Kawabata}, {Ohsugi}, {Yamashita}, {Isogai}, {Sato}, \&
  {Kino}}]{2008PASJ...60L..37S}
  {Sasada}, M., {Uemura}, M., {Arai}, A., {et~al.} 2008, \pasj, 60, L37

\bibitem[{{Schulz} \& {Lenzen}(1983)}]{1983A&A...121..158S}
{Schulz}, A., \& {Lenzen}, R. 1983, \aap, 121, 158

\bibitem[{{Stickel} {et~al.}(1991){Stickel}, {Padovani}, {Urry}, {Fried}, \&
  {Kuehr}}]{1991ApJ...374..431S}
{Stickel}, M., {Padovani}, P., {Urry}, C.~M., {Fried}, J.~W., \& {Kuehr}, H.
  1991, \apj, 374, 431

\bibitem[Tchekhovskoy \& Bromberg(2016)]{2016MNRAS.461L..46T}
  Tchekhovskoy, A., \& Bromberg, O.\ 2016, \mnras, 461, L46

\bibitem[{{Uemura} {et~al.}(2010){Uemura}, {Kawabata}, {Sasada}, {Ikejiri},
  {Sakimoto}, {Itoh}, {Yamanaka}, {Ohsugi}, {Sato}, \&
  {Kino}}]{2010PASJ...62...69U}
  {Uemura}, M., {Kawabata}, K.~S., {Sasada}, M., {et~al.} 2010, \pasj, 62, 69

\bibitem[{{Urry} \& {Padovani}(1995)}]{1995PASP..107..803U}
{Urry}, C.~M., \& {Padovani}, P. 1995, \pasp, 107, 803

\bibitem[{{Visvanathan} \& {Wills}(1998)}]{1998AJ....116.2119V}
{Visvanathan}, N., \& {Wills}, B.~J. 1998, \aj, 116, 2119

\bibitem[{{Watanabe} {et~al.}(2005){Watanabe}, {Nakaya}, {Yamamuro}, {Zenno},
  {Ishii}, {Okada}, {Yamazaki}, {Yamanaka}, {Kurita}, {Kino}, {Ijiri}, {Hirao},
  {Nagata}, {Sato}, {Kawai}, {Nakamura}, {Sato}, {Ebizuka}, {Hough}, \&
  {Chrysostomou}}]{2005PASP..117..870W}
{Watanabe}, M., {Nakaya}, H., {Yamamuro}, T., {et~al.} 2005, \pasp, 117, 870

\bibitem[{{Westfold}(1959)}]{1959ApJ...130..241W}
{Westfold}, K.~C. 1959, \apj, 130, 241

\bibitem[{{Wolff} {et~al.}(1996){Wolff}, {Nordsieck}, \&
  {Nook}}]{1996AJ....111..856W}
{Wolff}, M.~J., {Nordsieck}, K.~H., \& {Nook}, M.~A. 1996, \aj, 111, 856

\end{thebibliography}

\clearpage
\clearpage
\section*{Appendix A}

\begin{figure}
  \includegraphics[angle=0,width=8cm]{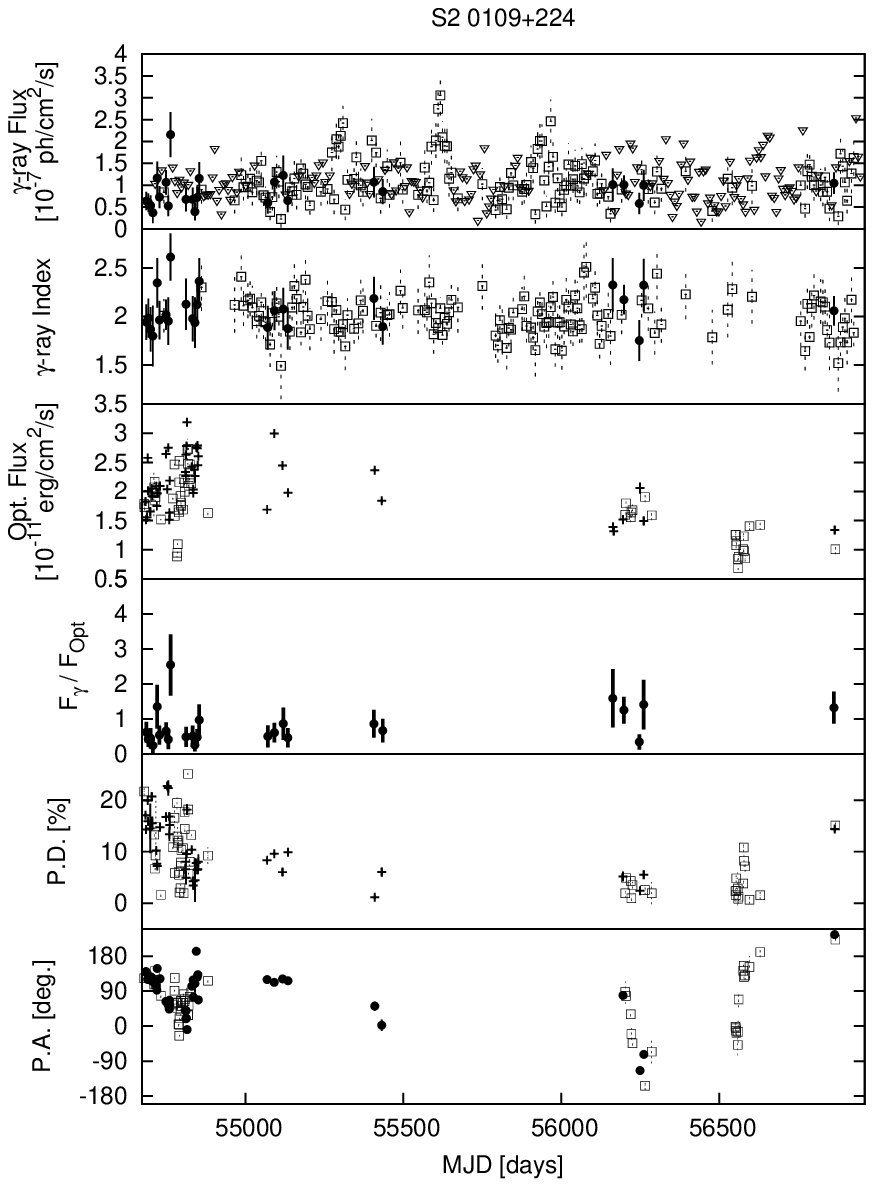}
  \caption{Multiwavelength light curves of S2 0109+224.
    Details of plots are shown in Fig. \ref{fig:MWLC_s5_0716}.
    Non-simultaneous data are represented by open boxes.
    Upper limits on the gamma-ray flux are indicated by open triangles.}
  \label{fig:LC_}
  \includegraphics[angle=0,width=8cm]{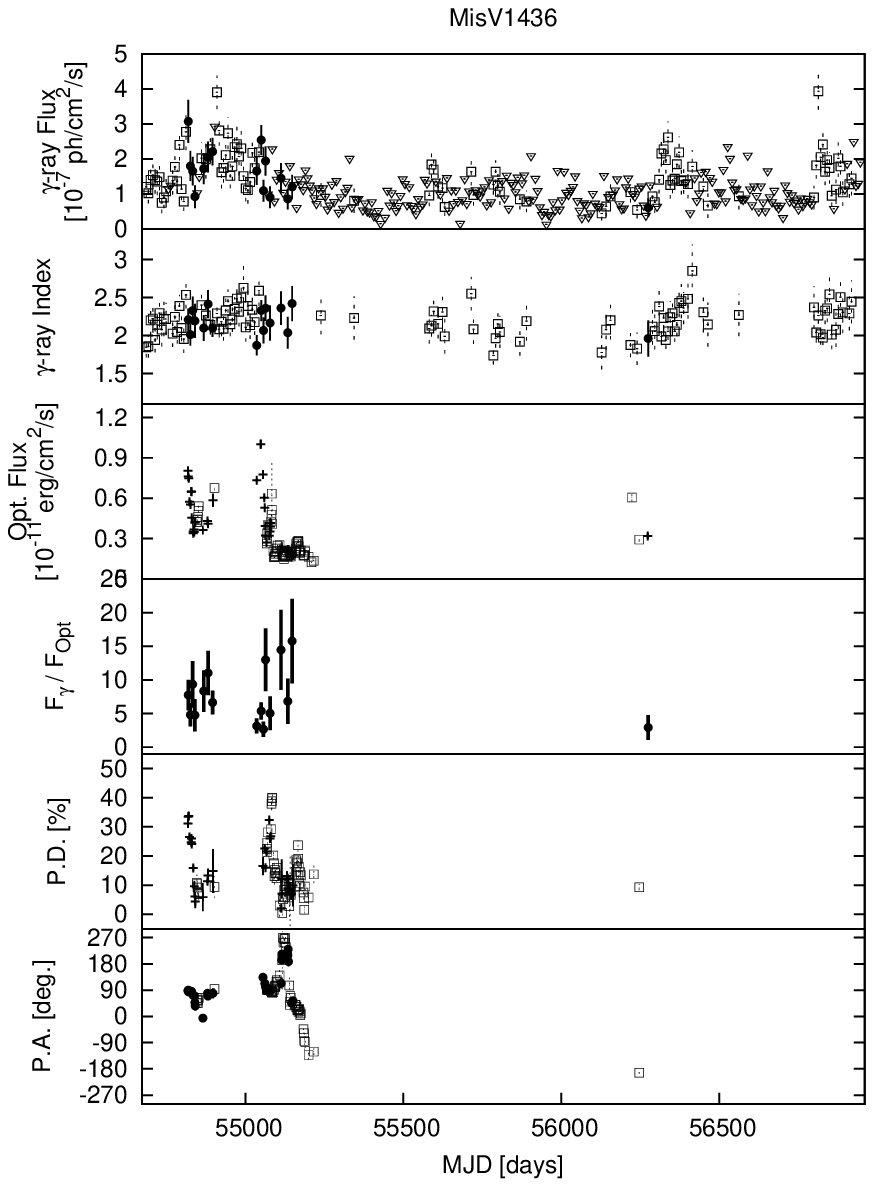}
  \caption{Multiwavelength light curves of MisV 1436}
  \label{fig:LC_}
\end{figure}
\begin{figure}
  \includegraphics[angle=0,width=8cm]{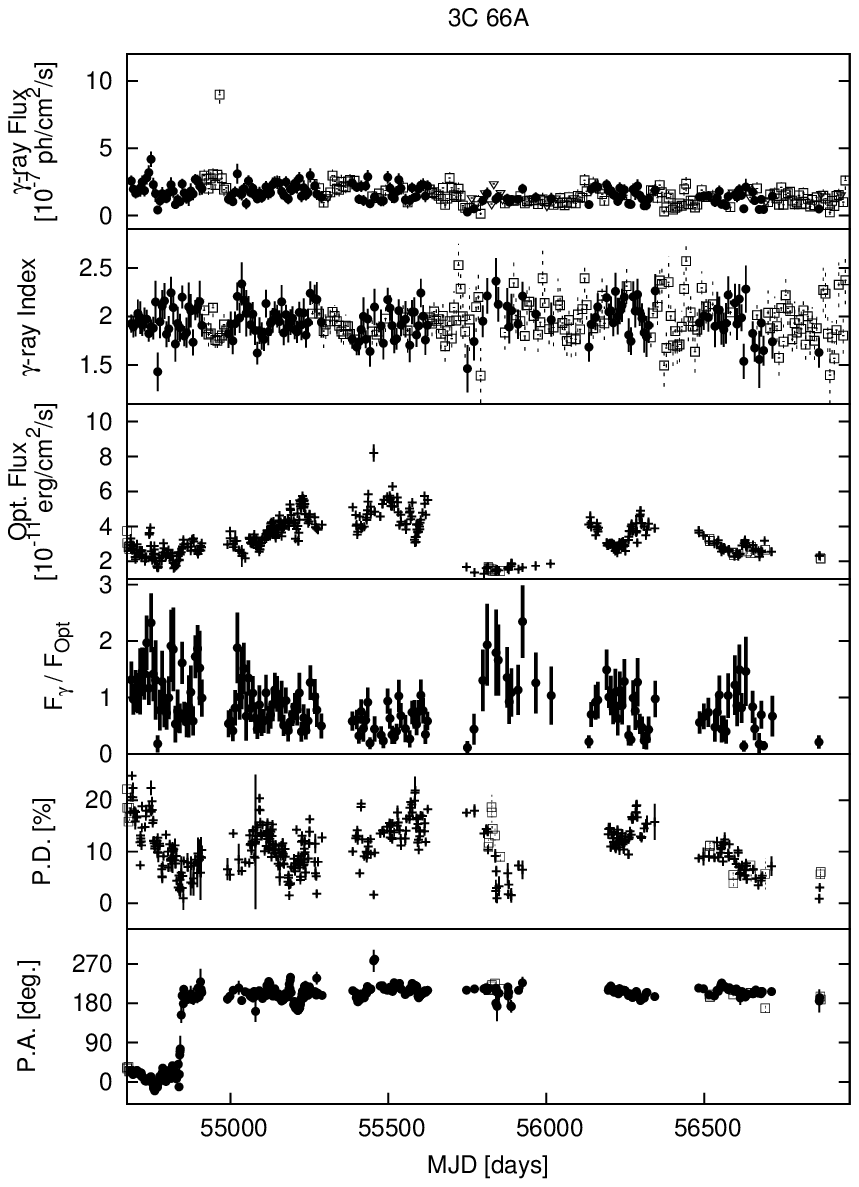}
  \caption{Multiwavelength light curves of 3C 66A}
  \label{fig:LC_3C 66A}
  \includegraphics[angle=0,width=8cm]{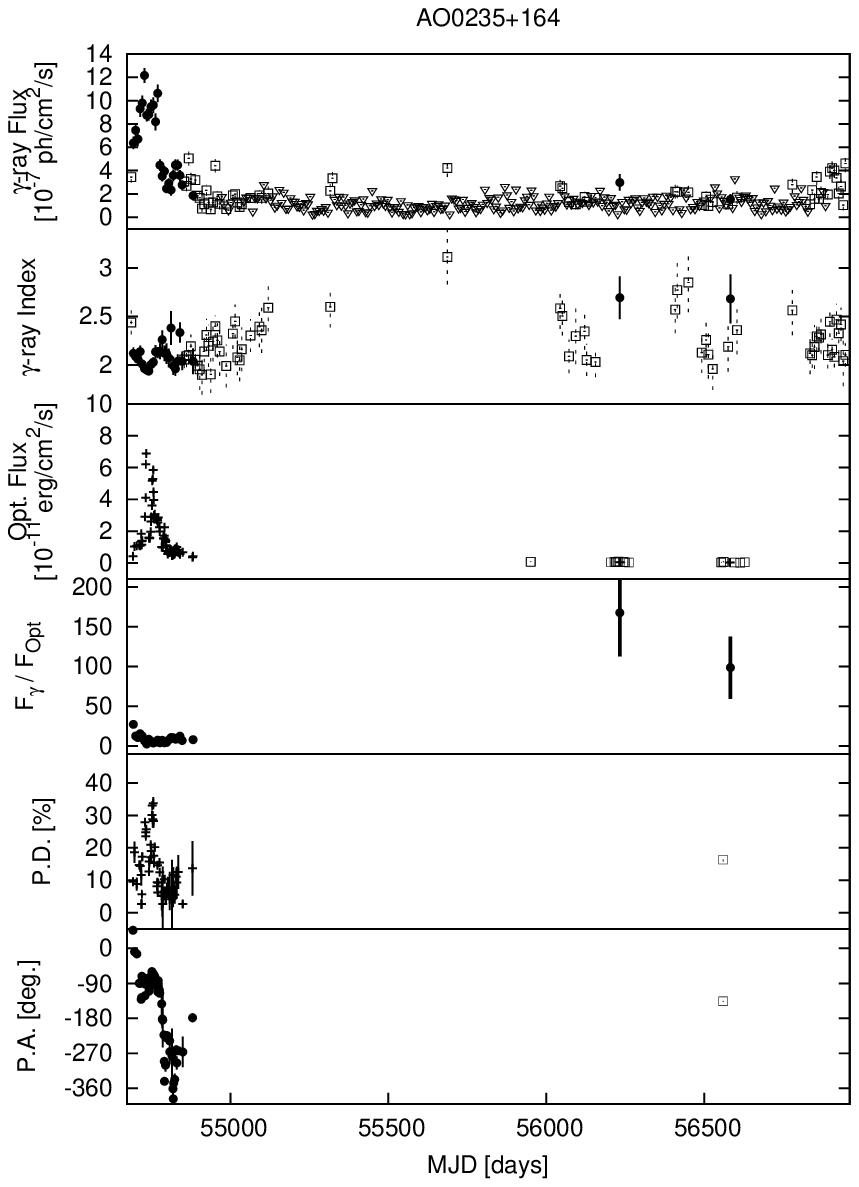}
  \caption{Multiwavelength light curves of AO 0235+164}
  \label{fig:LC_}
\end{figure}

\clearpage
\begin{figure}
  \includegraphics[angle=0,width=8cm]{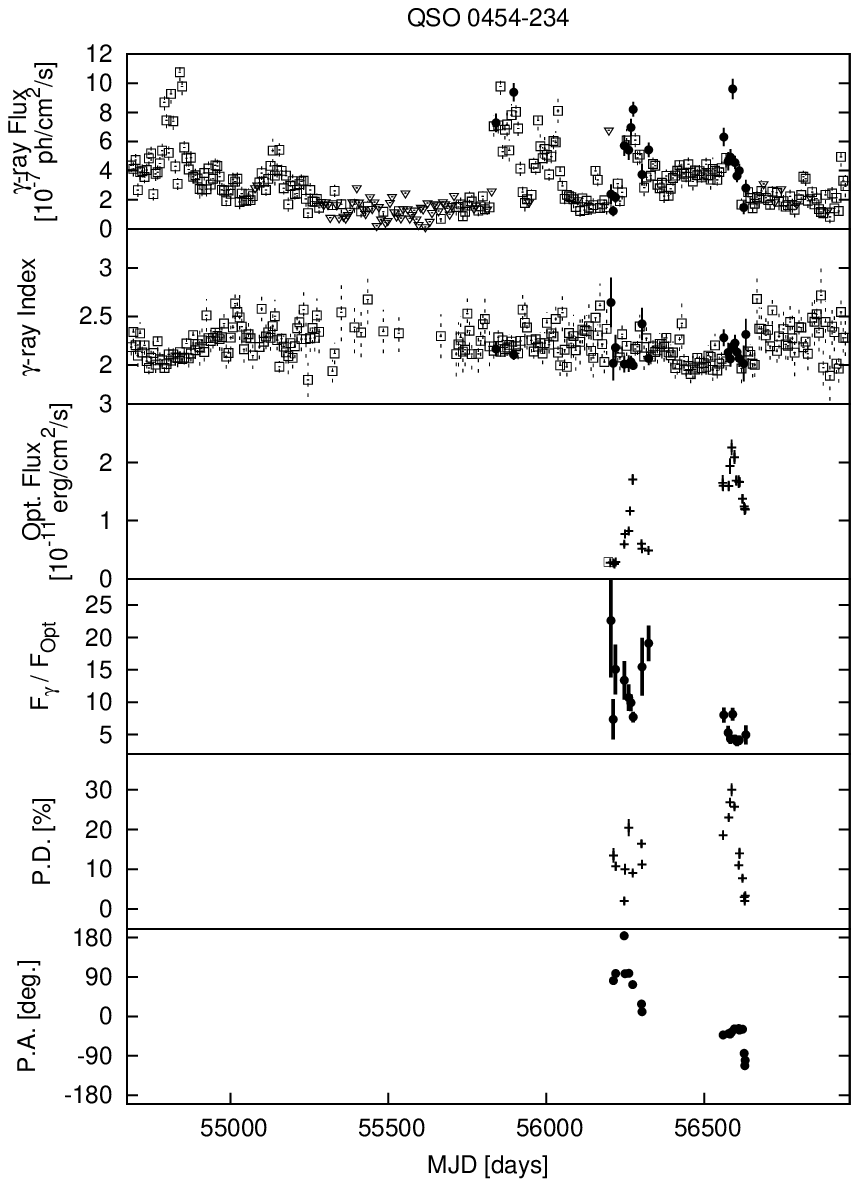}
  \caption{Multiwavelength light curves of PKS 0454+234}
  \label{fig:LC_}
  \includegraphics[angle=0,width=8cm]{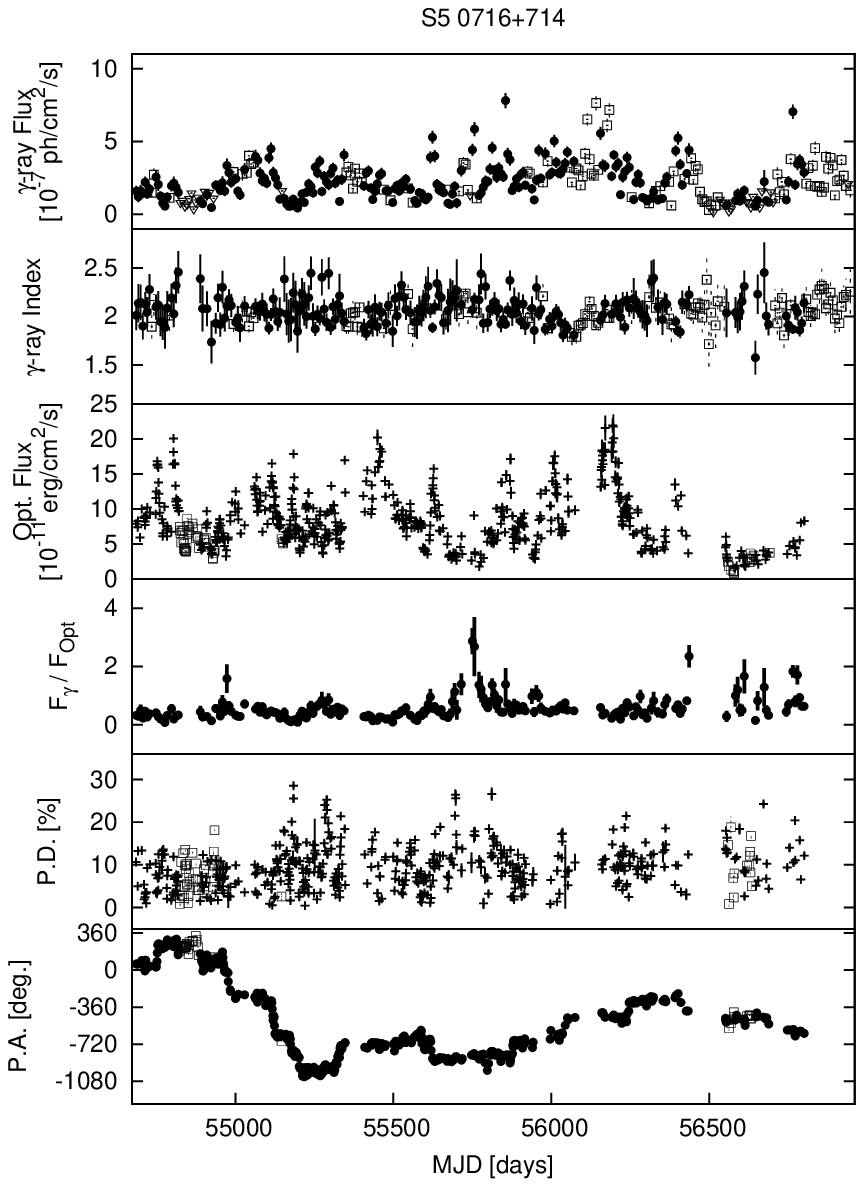}
  \caption{Multiwavelength light curves of S5 0716+714}
  \label{fig:LC_S5_2}
\end{figure}
\begin{figure}
  \includegraphics[angle=0,width=8cm]{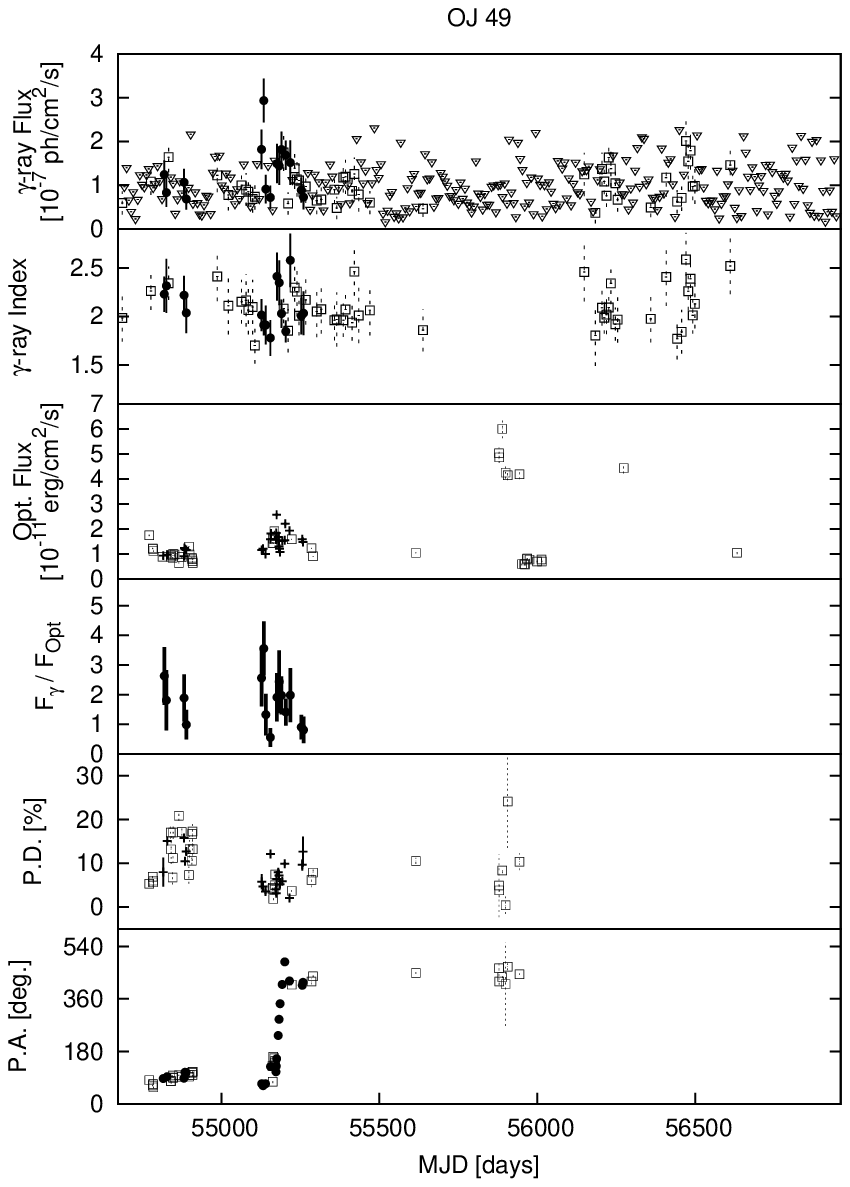}
  \caption{Multiwavelength light curves of OJ 49}
  \label{fig:LC_}
  \includegraphics[angle=0,width=8cm]{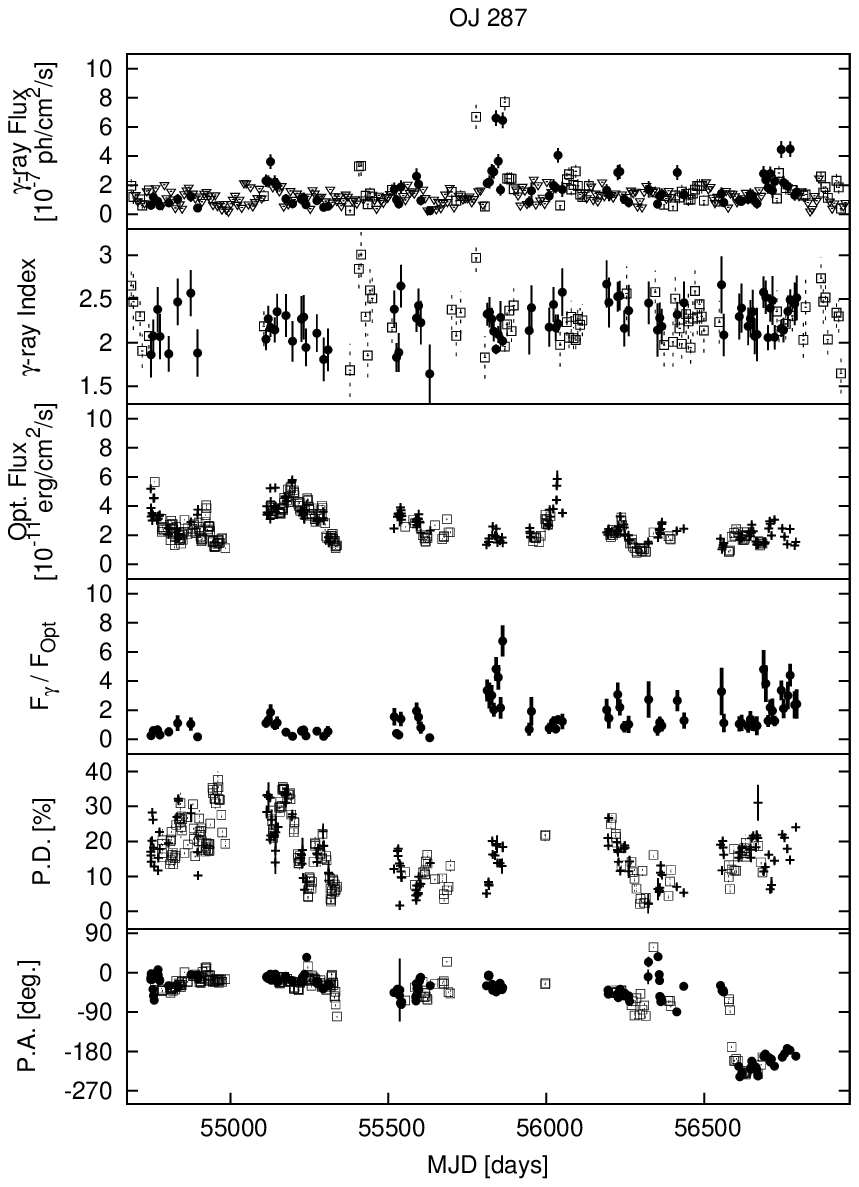}
  \caption{Multiwavelength light curves of OJ 287}
  \label{fig:LC_}
\end{figure}

\clearpage
\begin{figure}
  \includegraphics[angle=0,width=8cm]{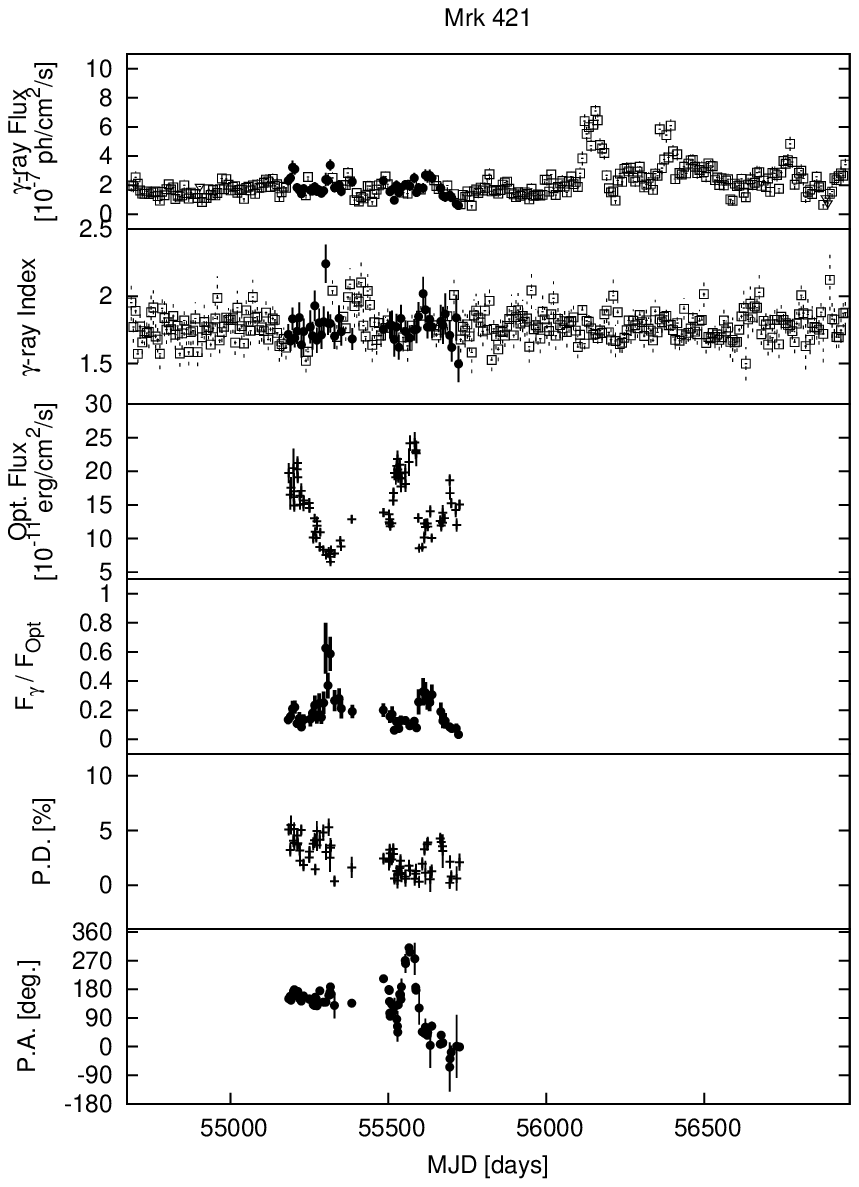}
  \caption{Multiwavelength light curves of Mrk 421}
  \label{fig:LC_}
  \includegraphics[angle=0,width=8cm]{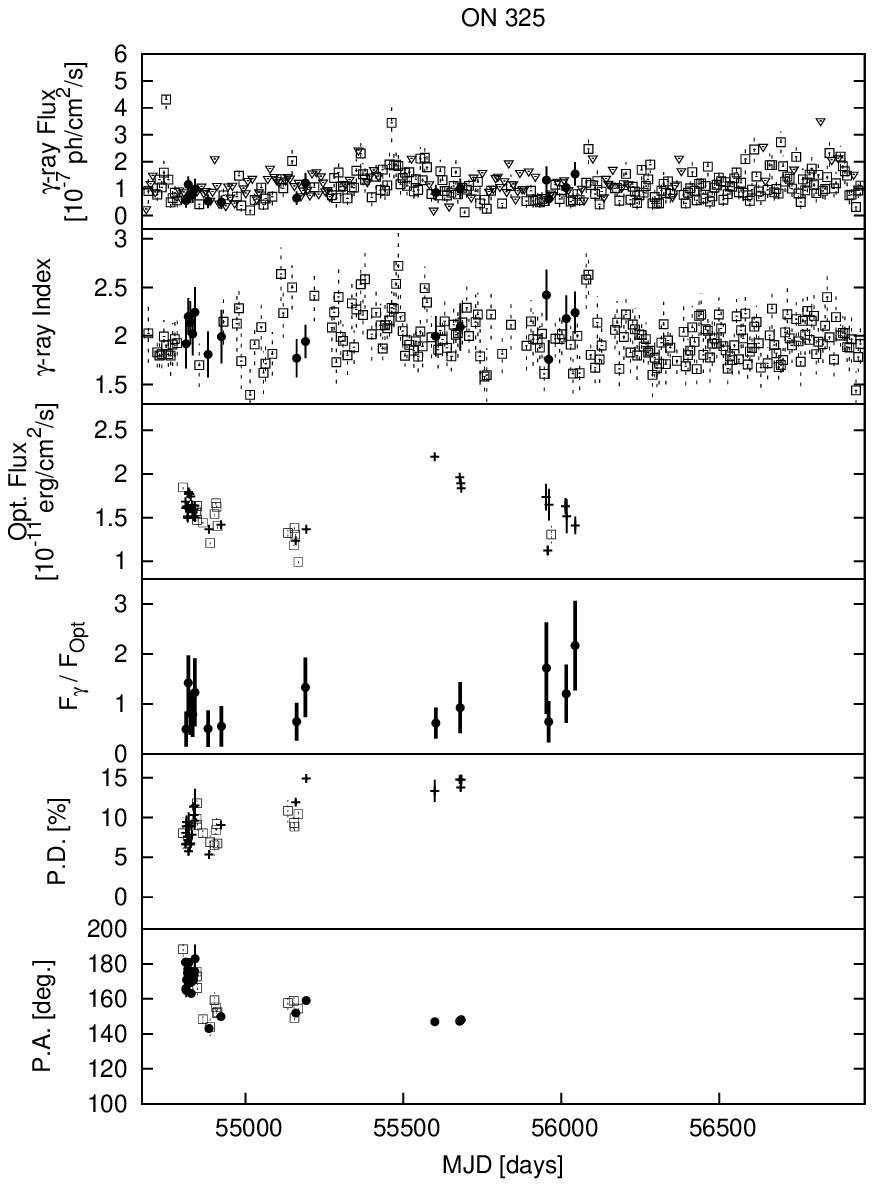}
  \caption{Multiwavelength light curves of ON 325}
  \label{fig:LC_}
\end{figure}
\begin{figure}
  \includegraphics[angle=0,width=8cm]{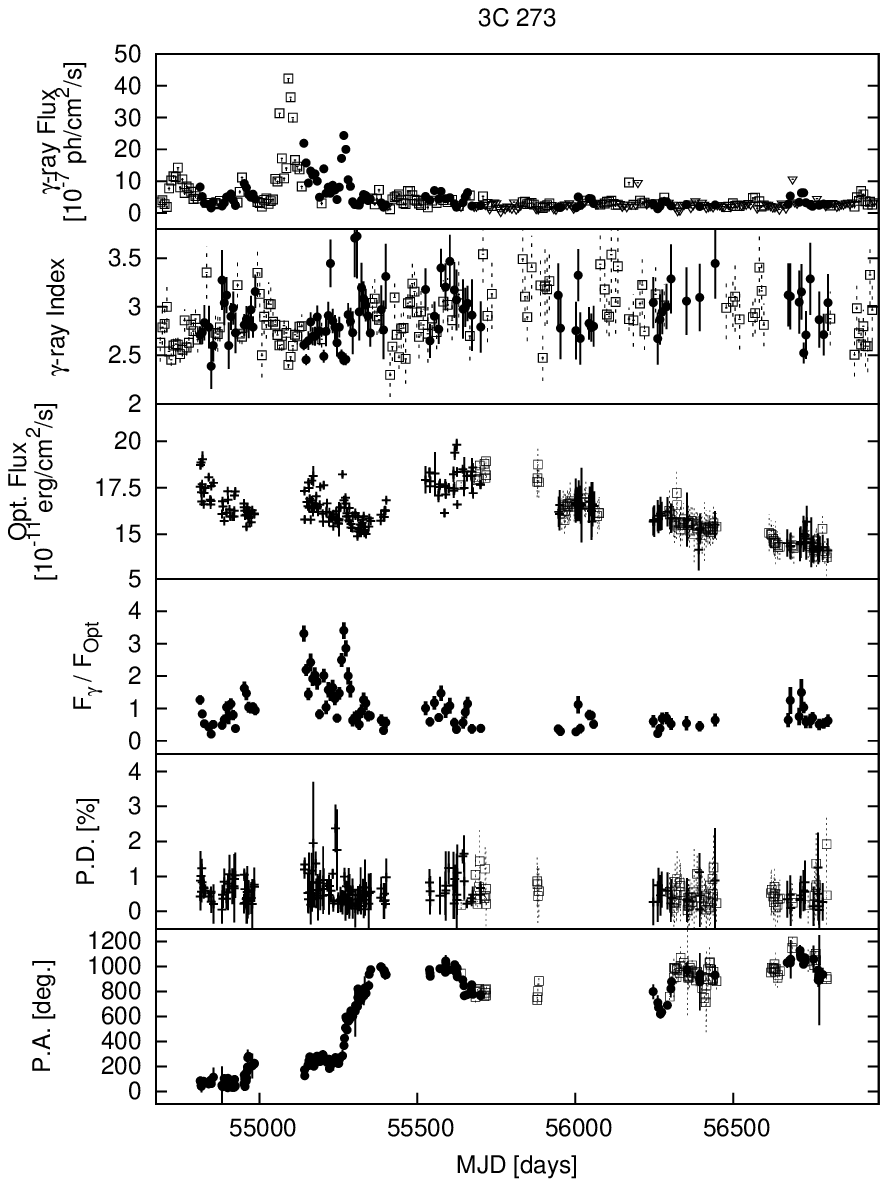}
  \caption{Multiwavelength light curves of 3C 273}
  \label{fig:LC_}
  \includegraphics[angle=0,width=8cm]{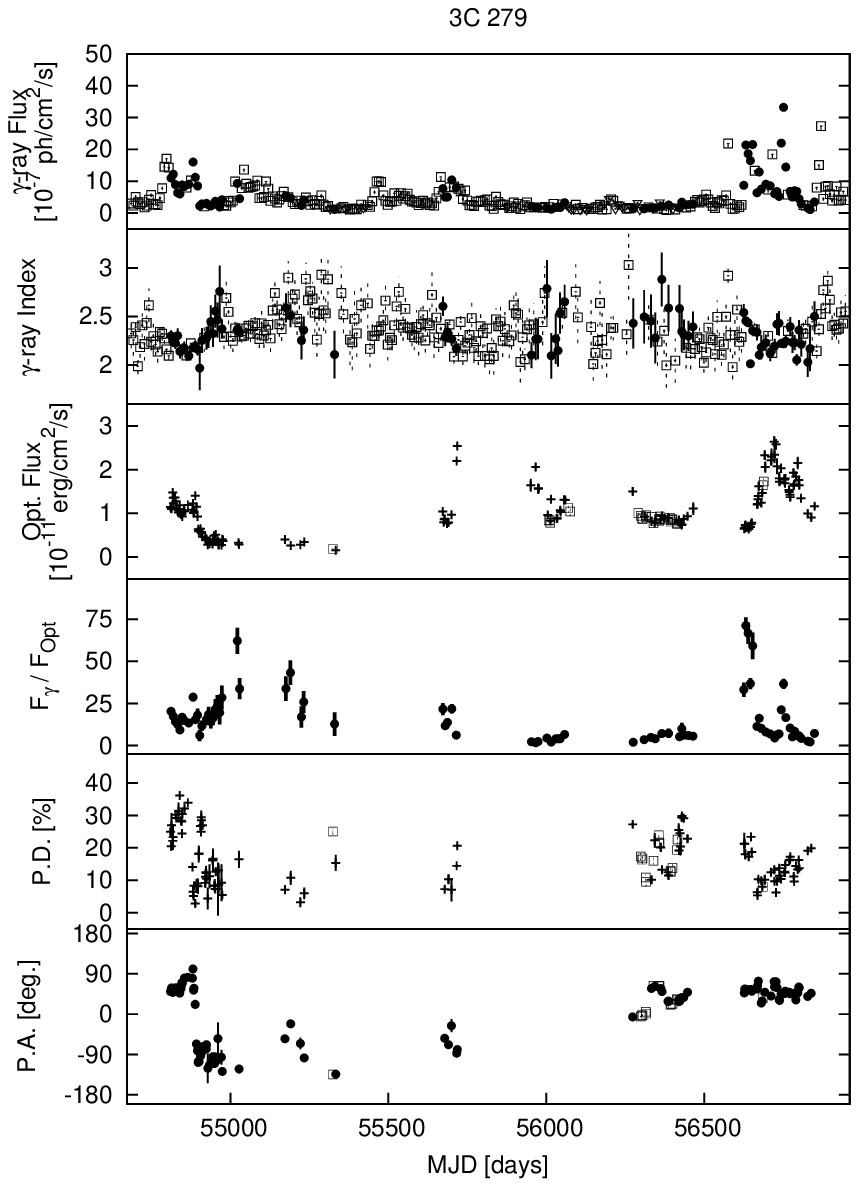}
  \caption{Multiwavelength light curves of 3C 279}
  \label{fig:LC_}
\end{figure}

\clearpage
\begin{figure}
  \includegraphics[angle=0,width=8cm]{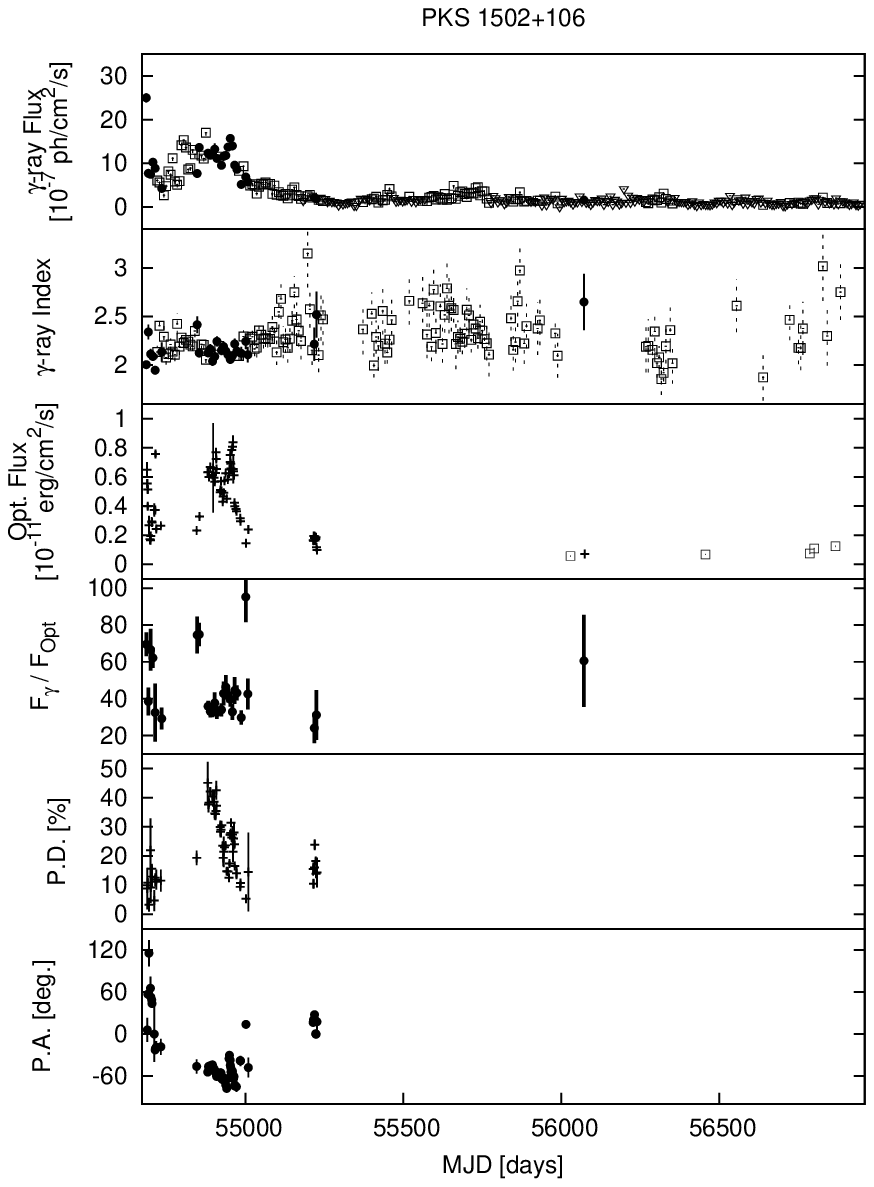}
  \caption{Multiwavelength light curves of PKS 1502+106}
  \label{fig:LC_}
  \includegraphics[angle=0,width=8cm]{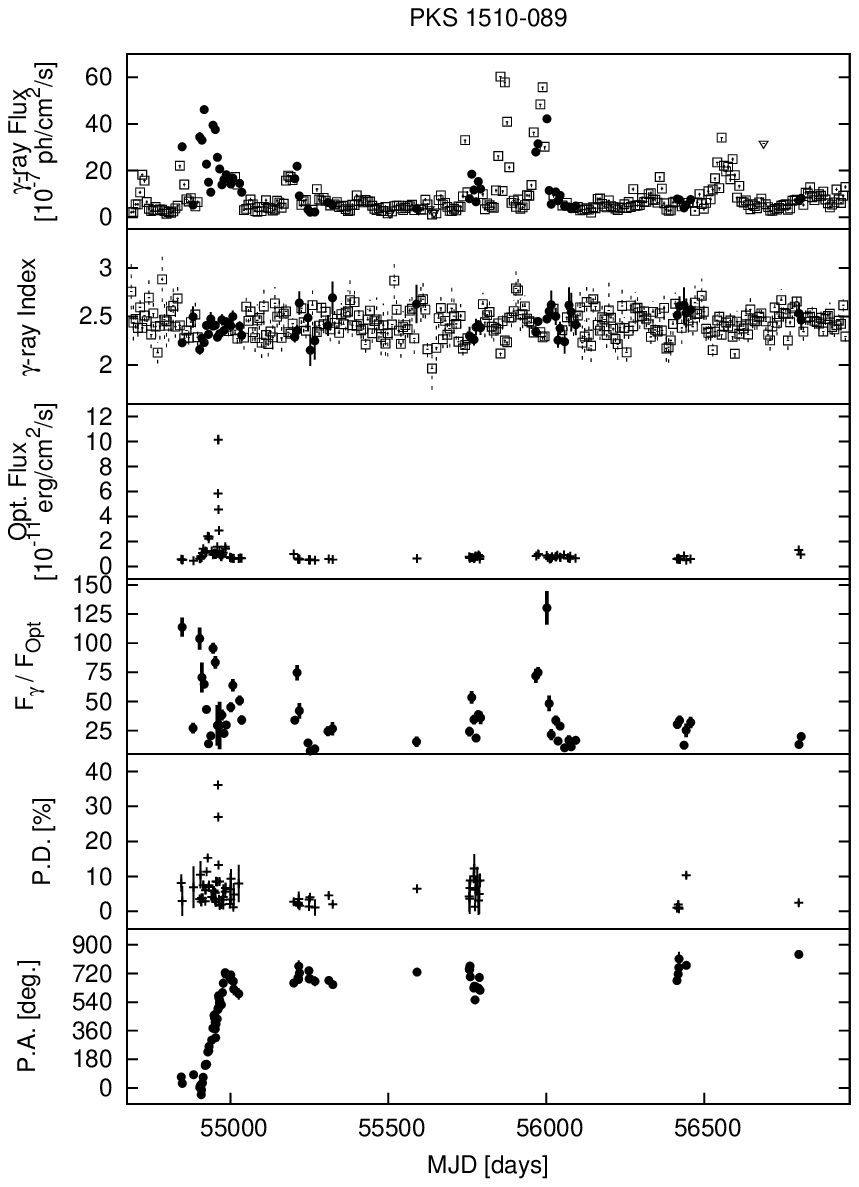}
  \caption{Multiwavelength light curves of PKS 1510-089}
  \label{fig:LC_}
\end{figure}
\begin{figure}
  \includegraphics[angle=0,width=8cm]{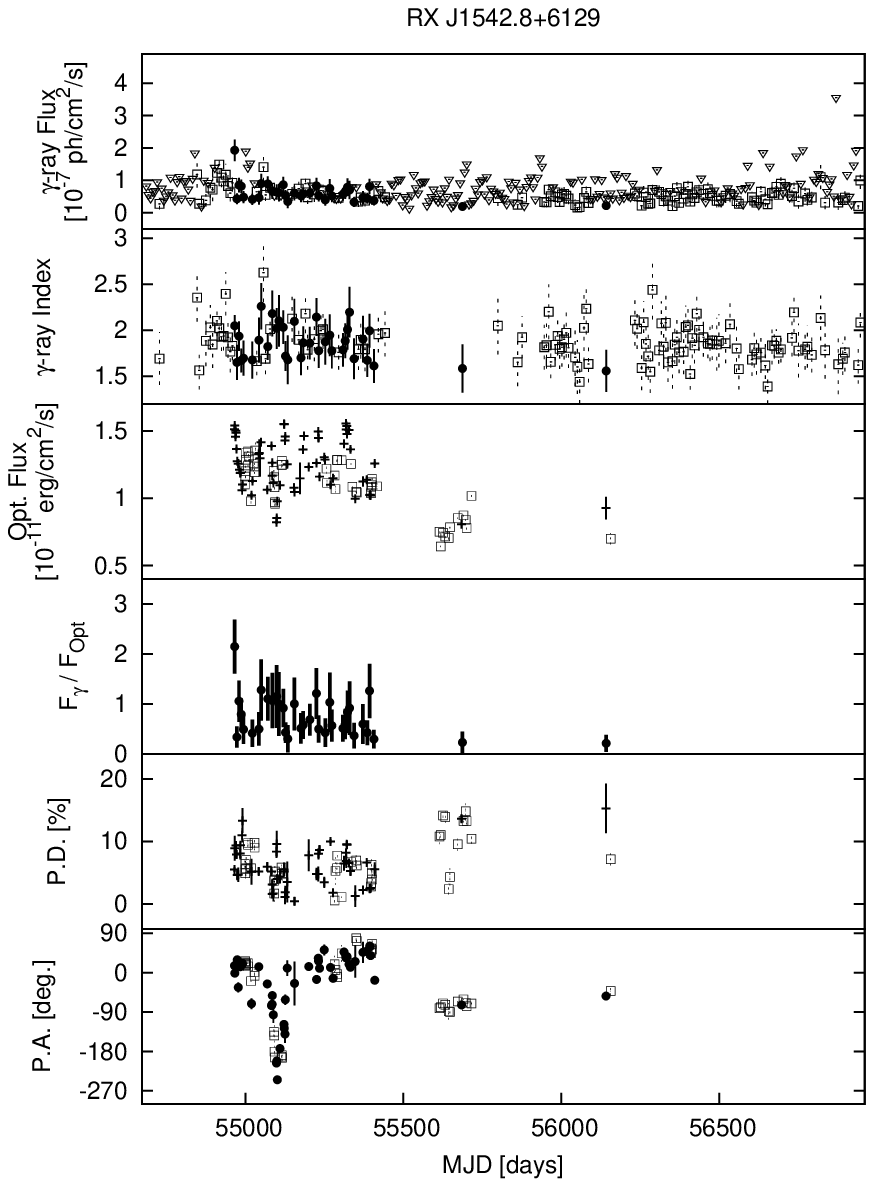}
  \caption{Multiwavelength light curves of RX J1542.8+6129}
  \label{fig:LC_}
  \includegraphics[angle=0,width=8cm]{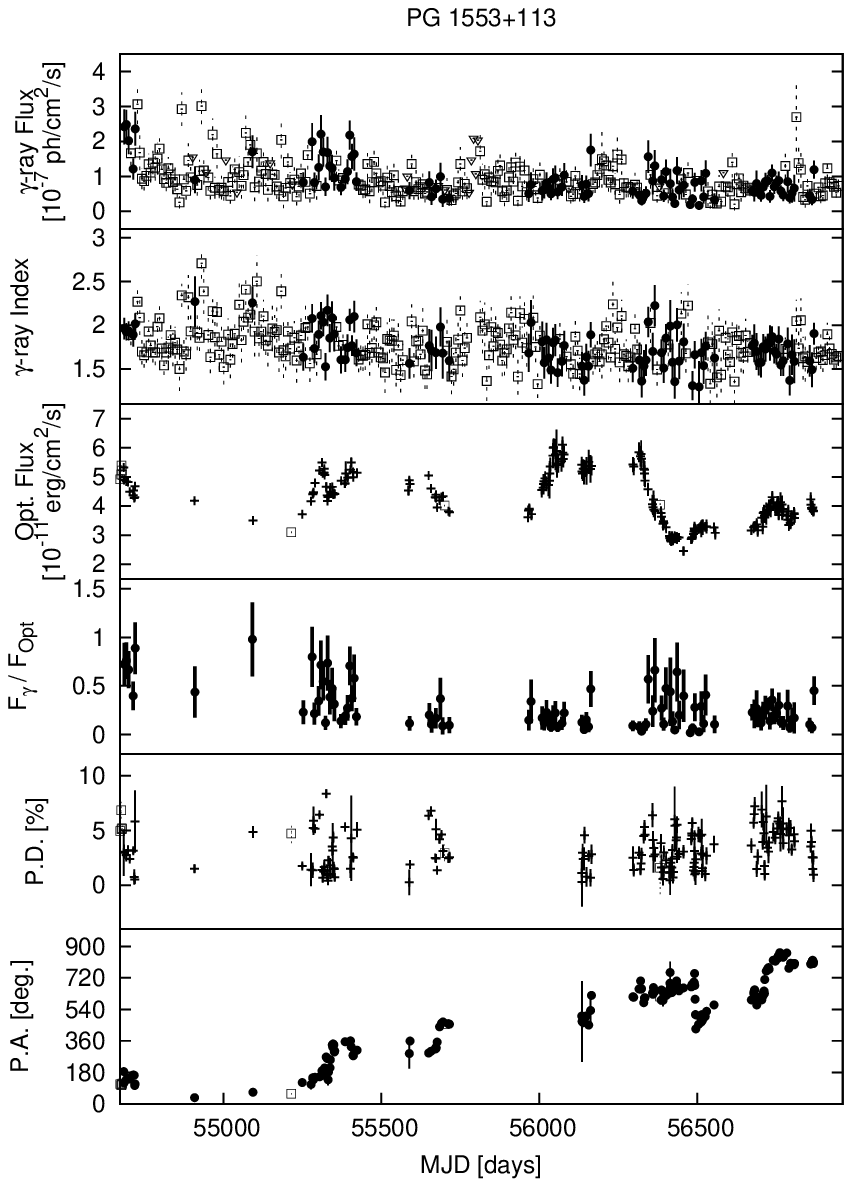}
  \caption{Multiwavelength light curves of PG 1553+113}
  \label{fig:LC_}
\end{figure}

\clearpage
\begin{figure}
  \includegraphics[angle=0,width=8cm]{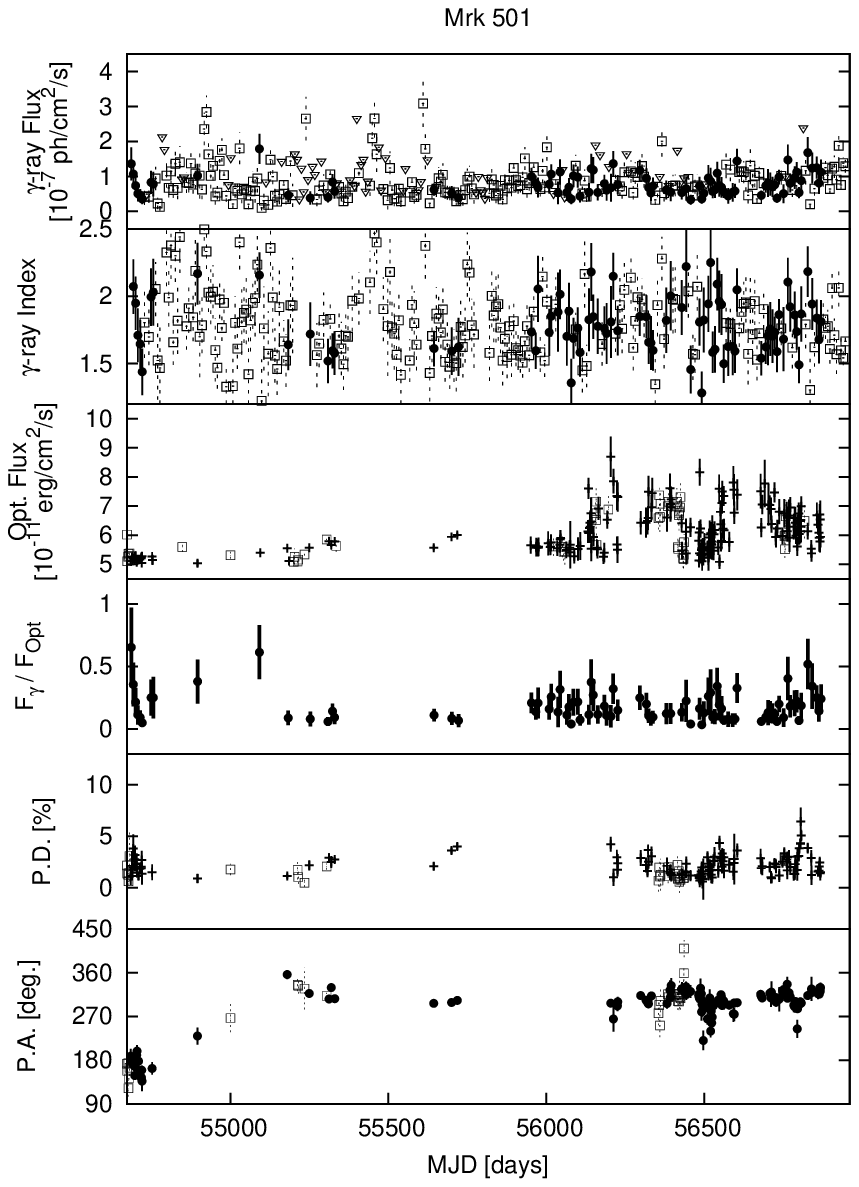}
  \caption{Multiwavelength light curves of Mrk 501}
  \label{fig:LC_}
  \includegraphics[angle=0,width=8cm]{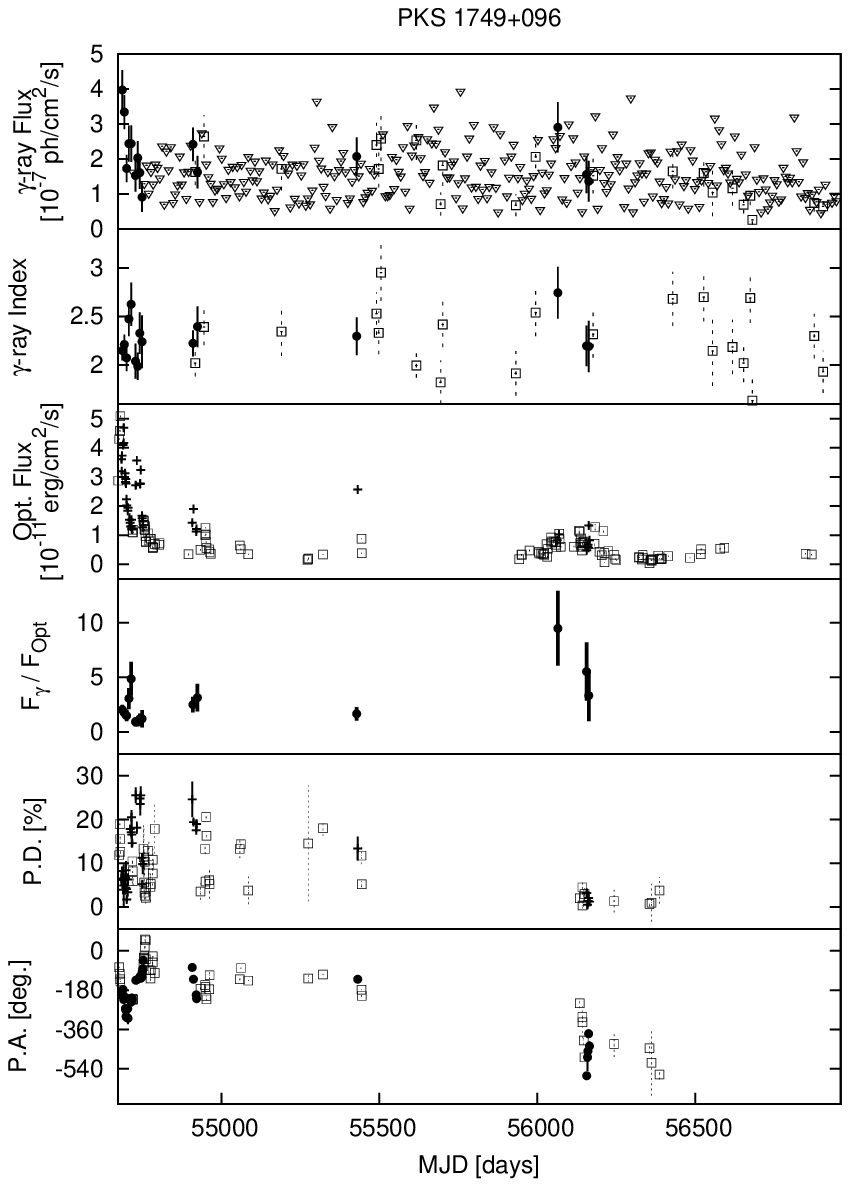}
  \caption{Multiwavelength light curves of PKS 1749+096}
  \label{fig:LC_}
\end{figure}
\begin{figure}
  \includegraphics[angle=0,width=8cm]{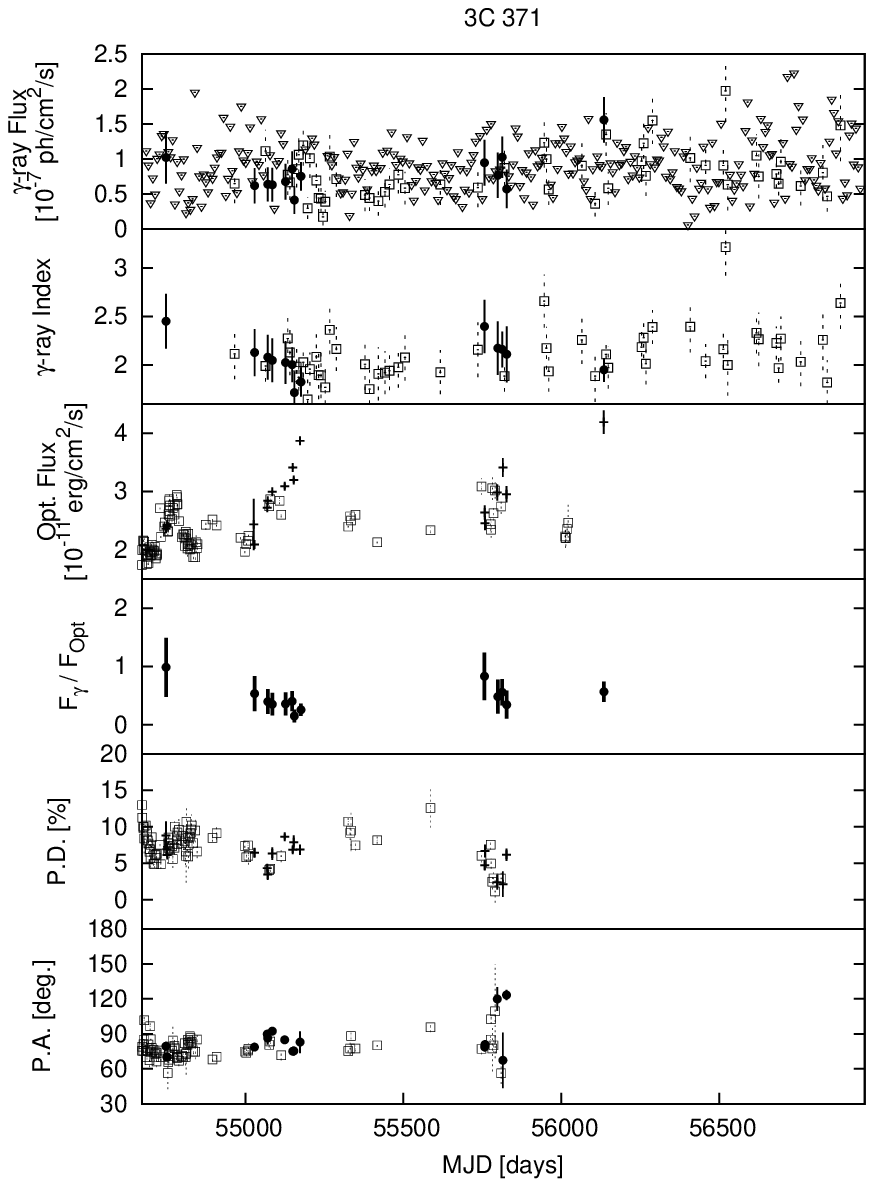}
  \caption{Multiwavelength light curves of 3C 371}
  \label{fig:LC_}
  \includegraphics[angle=0,width=8cm]{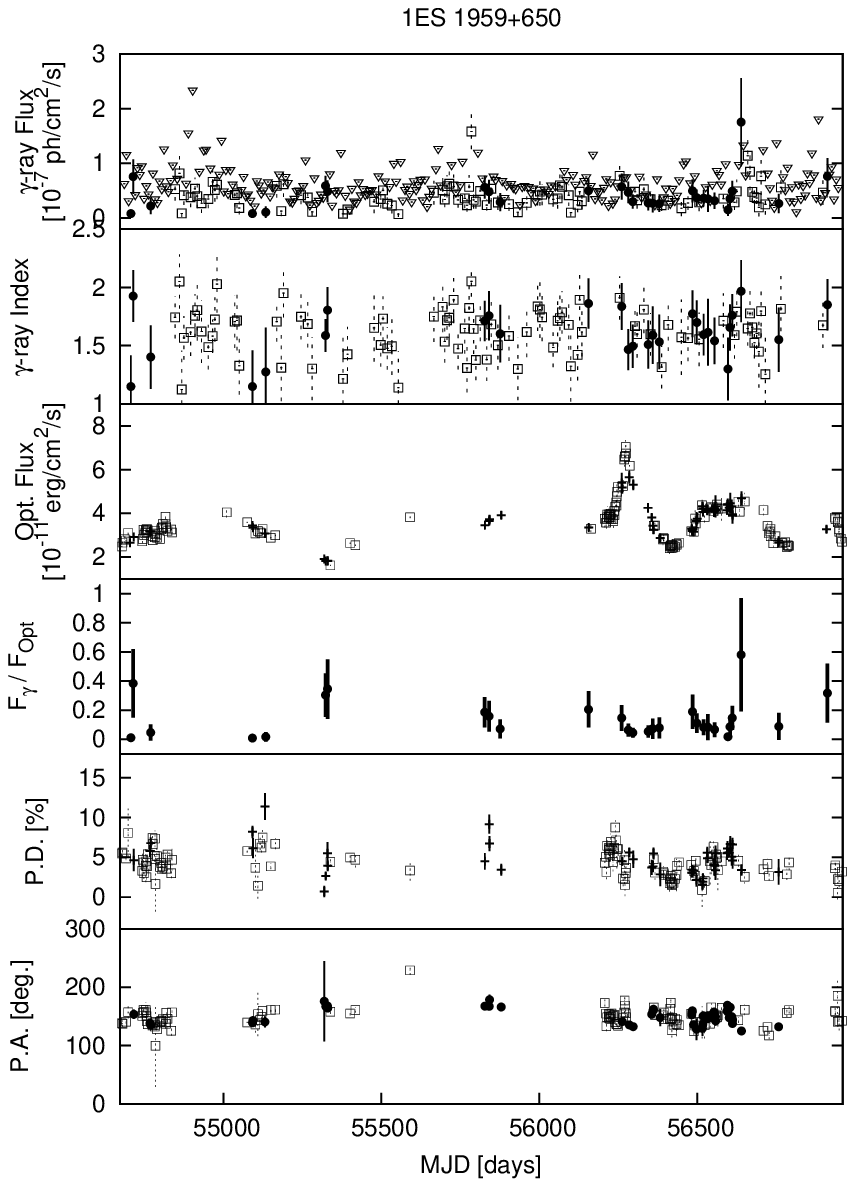}
  \caption{Multiwavelength light curves of 1ES 1959+650}
  \label{fig:LC_}
\end{figure}

\clearpage
\begin{figure}
  \includegraphics[angle=0,width=8cm]{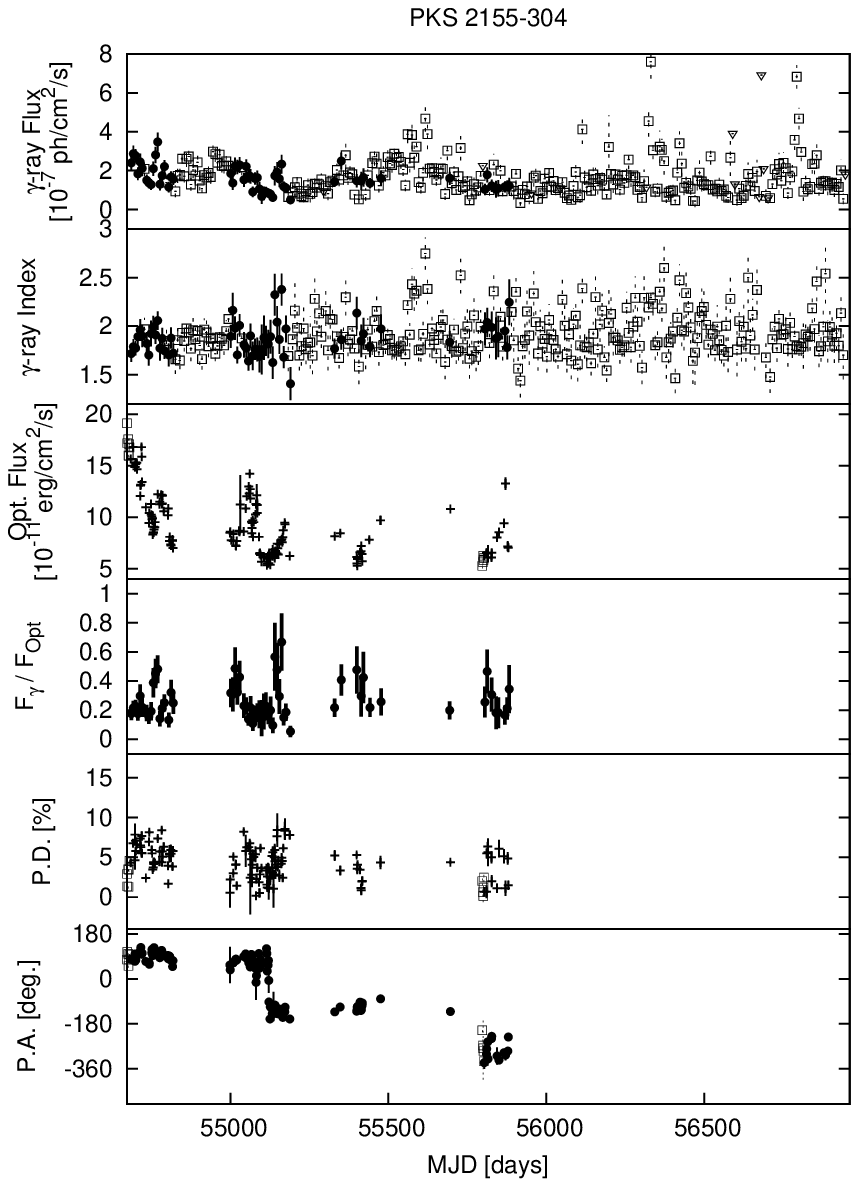}
  \caption{Multiwavelength light curves of PKS 2155-304}
  \label{fig:LC_}
  \includegraphics[angle=0,width=8cm]{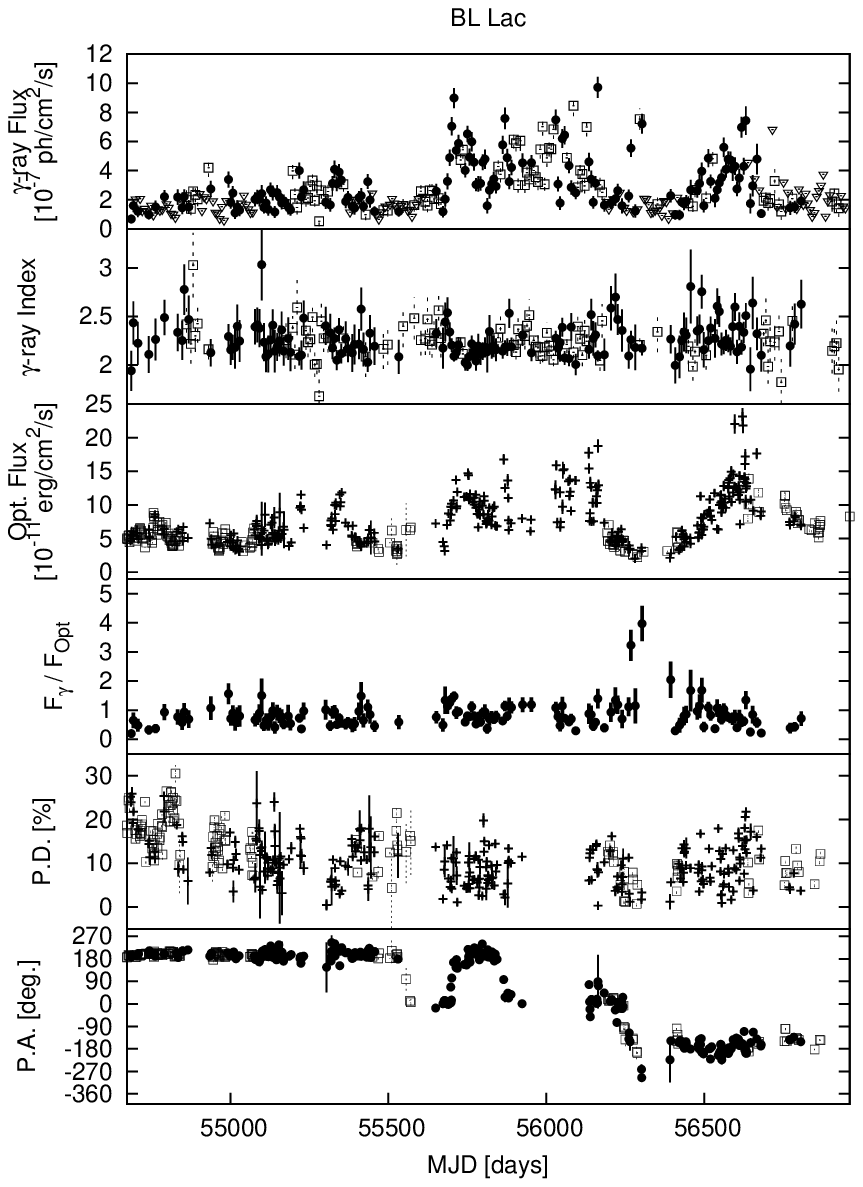}
  \caption{Multiwavelength light curves of BL Lac}
  \label{fig:LC_}
\end{figure}
\begin{figure}
  \includegraphics[angle=0,width=8cm]{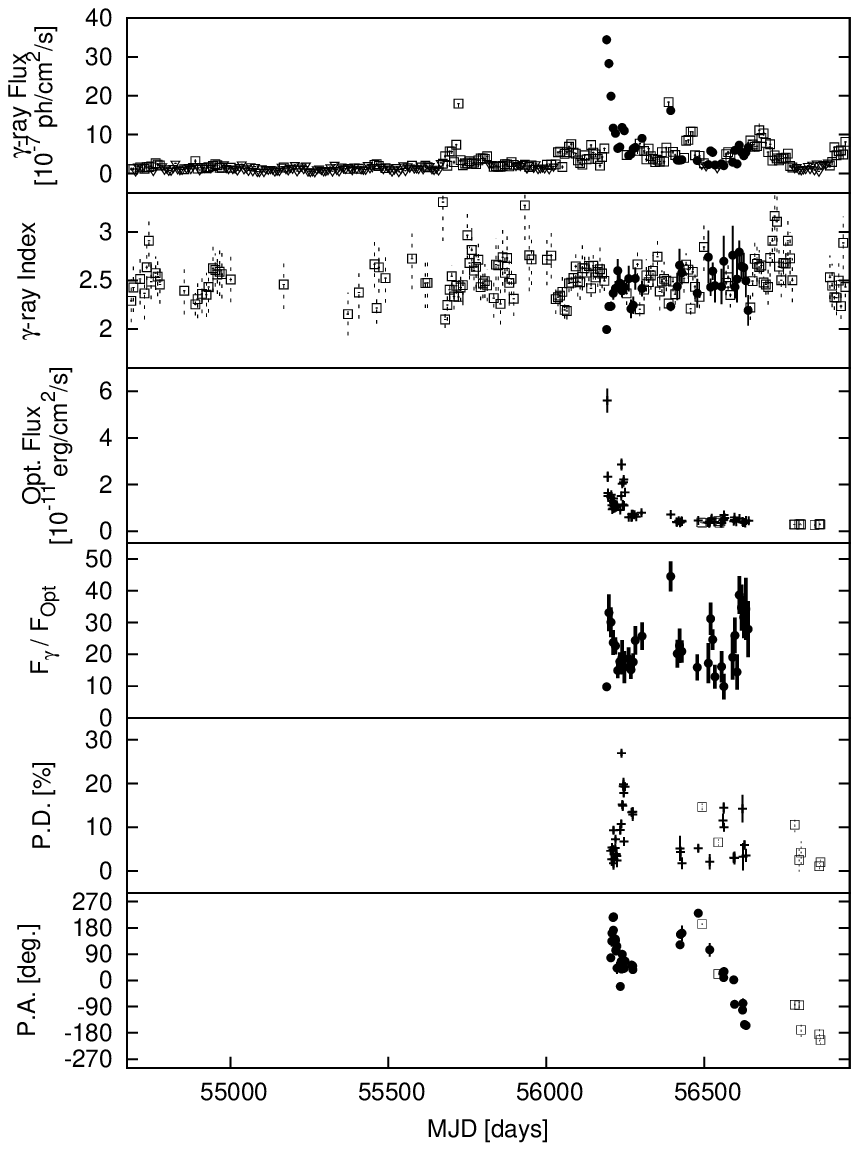}
  \caption{Multiwavelength light curves of CTA 102}
  \label{fig:LC_}
  \includegraphics[angle=0,width=8cm]{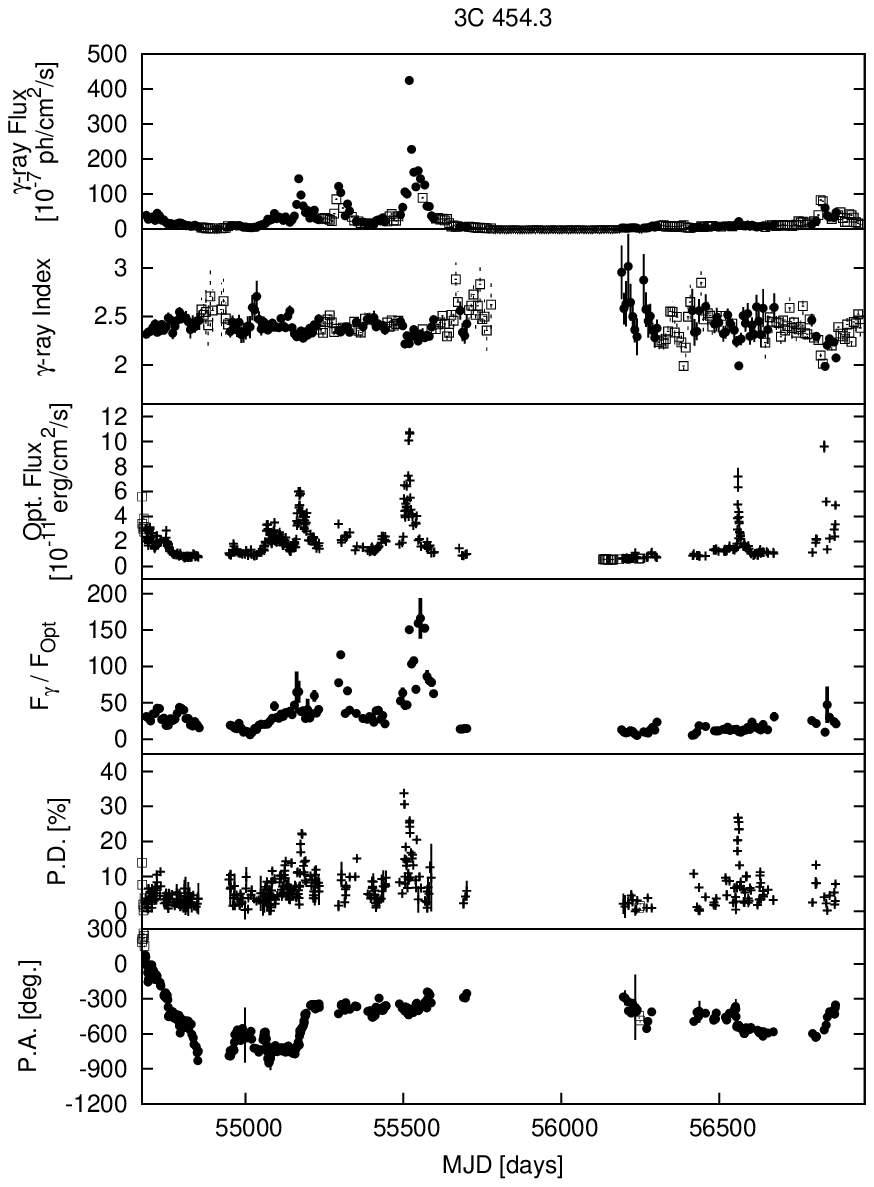}
  \caption{Multiwavelength light curves of 3C 454.3}
  \label{fig:LC_}
\end{figure}

\clearpage
\section*{Appendix B}

Basic block Bootstrap is a simulation method used to estimate the
distribution of test statistics. It is used for time series
data \citep[e.g.,][]{2008ApJ...681..726L}.
The original dataset is split into $N_{\rm block}$ non-overlapping blocks.
We then randomly resampled the dataset based on these blocks from
the original data, and calculated the correlation coefficient
between the original and the replacement data.
This routine was repeated 10,000 times to obtain the Bootstrap
distribution of correlation coefficient for each dataset.
From this Bootstrap distribution, a confidence interval of
$\alpha=0.95$ was derived.
For blazars, the typical timescale of a flare is about a few
weeks and the cadence of our dataset is typically a few days.
Therefore, $N_{\rm block} \sim 5$ corresponds to the typical timescale
of blazar flares.
In this paper, we fixed the $N_{\rm block}$ = 5 for all of our samples.
Of course, different blazars have different timescales but
we also confirmed that the confidence interval dependence
on block size is negligible.
Table \ref{tab:boot} summarizes the confidence level dependence on block
size for the correlation coefficient of gamma-ray flux and
optical flux with a time lag of zero for S5~0716+714.

\begin{table}[!htb]
  \begin{center}
  \caption{Confidence interval dependence on block size}
  \label{tab:boot} 
  \begin{tabular}{cc}\hline\hline
    Block size ($N_{\rm block}$) & 95\% C.I. \\ \hline
    3   & (0.1852  0.4846) \\
    4   & (0.1871, 0.4847) \\
    5   & (0.1887, 0.4842) \\
    6   & (0.1903, 0.4824) \\
    7   & (0.1885, 0.4826) \\
    8   & (0.1869, 0.4842) \\
    9   & (0.1885, 0.4831) \\
    10  & (0.1894, 0.4828) \\
    20  & (0.1904, 0.4826) \\
    30  & (0.1891, 0.4845) \\
    50  & (0.1922, 0.4843) \\
    100 & (0.1960, 0.4881) \\
\hline
  \end{tabular}
  \end{center}
\end{table}

\end{document}